\documentclass[]{jfm}

\usepackage{graphicx}
\usepackage{epstopdf,epsfig}
\usepackage{newtxtext}
\usepackage{newtxmath}
\usepackage{natbib}
\usepackage{hyperref}
\usepackage{caption}
\usepackage{subcaption}

\hypersetup{
    colorlinks = true,
    urlcolor   = blue,
    citecolor  = black,
}

\newcommand{\RomanNumeralCaps}[1]
\shorttitle{Optimal projections.}
\shortauthor{A. Pradhan, and K. Duraisamy}

\newcommand{\comment}[1]{}

\title{A unified understanding of scale-resolving simulations and near-wall modeling of turbulent flows using optimal finite element projections.}

\author{Aniruddhe Pradhan\aff{1}
  \corresp{\email{anipra@umich.edu}},
  Karthik Duraisamy\aff{1}}

\affiliation{\aff{1}Department of Aerospace Engineering, University of Michigan, Ann Arbor, MI 48109}

\begin{document}

\maketitle
 
\begin{abstract}
The main objective of this work is to develop a unified framework that can be used as a lens to quantitatively assess and augment a wide range of coarse-grained models of turbulence, viz. large eddy simulations (LES), hybrid Reynolds-averaged/LES methods and wall-modeled (WM)LES. Taking a turbulent channel flow as an example, optimality is assessed in the wall-resolved limit, the hybrid RANS/LES limit and the WMLES limit, via projections at different resolutions suitable for these approaches.  These optimal {\color{black} a priori} estimates are shown to have  similar characteristics to existing  {\color{black} a posteriori} solutions reported in the literature. Consistent accuracy metrics are developed for scale-resolving methods using the optimal solution as a reference, and evaluations are performed. We further characterize the slip-velocity in WMLES in terms of the near-wall under-resolution and  develop a universal scaling relationship.  {\color{black}Insights from the a-priori tests are used to augment  existing slip-based wall models. Various a posteriori tests reveal  superior performance over the dynamic slip wall model}. Guidance for the development of improved slip-wall models is provided, including a target for the dynamic procedure.  
\end{abstract}


\section{Introduction}
{\color{black}Simulation of turbulent flows  remains a challenge because of the disparate range of spatial and temporal scales that need to be resolved~\citep{pope2000turbulent}. An alternate to directly solving the Navier-Stokes equations is to solve its reduced complexity versions.  Reynolds averaged Navier-Stokes (RANS) models solve for the ensemble average or time-average of the true solution. Large Eddy Simulations (LES) \citep{DSM,GDSM,DSM2,SIGMA,VREMEN,WALE,OSS,OSS2,VMS3,VMSE,NLVMS,NLVMS2,MZVMS}  resolve the spatio-temporal dynamics of the large scales. The cost of LES is, however, still prohibitive near the wall. To alleviate the need of mesh refinement near the wall, boundary conditions are imposed weakly  in a wall-modelled LES (WMLES) \citep{piomelli2002wall,bose2014dynamic,bae2019dynamic}.  Alternate approaches to WMLES are hybrid RANS-LES techniques like DES \citep{spalart2009detached} and IDDES \citep{shur1999detached}, where the inner-layer is solved using RANS and the rest using LES. As a consequence, the cost associated with resolving the near-wall structures in the stream-wise and span-wise direction is no longer present. The cost associated with resolving the wall-normal gradient is still present in  the hybrid RANS-LES approaches. 

Over the past few decades, various contributions have been made in the development and application of these methods to highly complex problems (e.g. \cite{goc2020wall,lozano2020prediction,park2016wall,iyer2020wall,goc2021large,kiris2022high}). Our view is that, since all of the scale-resolving methods are coarse-grained from the Navier--Stokes equations, there must exist a unified view. The Partially-averaged Navier Stokes (PANS) approach brings together several turbulence closures of various modeled-to-resolved scale ratios ranging from Reynolds-averaged Navier Stokes (RANS) to Navier-Stokes (direct numerical simulations (DNS)) into one formulation \citep{girimaji2005partially}. The behavior of the PANS equations can be varied smoothly from the RANS equations to the Navier-Stokes (DNS) equations by changing the filter-width control parameters.  The unified RANS-LES approach~\cite{heinz2007unified,gopalan2013unified} is an optimal hybrid RANS–LES framework that uses different time-scales to switch between the RANS and LES approaches. In pursuit of similar unified models and in an effort to augment existing frameworks, we propose a filtering technique using optimal finite element projections which: (i) offers a unifying perspective through a common coarse-graining strategy; (ii) provides optimal solutions for the existing coarse-grained methods to improve upon.

The use of filtered DNS data to perform a priori analysis of closure models for RANS and LES is indeed not new.  LES models such as the scale-similarity or Smagorinsky models have also been frequently evaluated against sub-grid stresses obtained from filtered DNS data \citep{vreman1995priori,bou2008scale,meneveau2000scale,girimaji2005partially}. In most of the prior studies, filtering is either performed in the Fourier space using the sharp spectral cutoff or Gaussian filters when the problem has periodic directions or the box-filter in more complex problems. In case of filters that are applied in  spectral space, the filter width remains the same along the periodic directions in which it is applied. However, as observed in most coarse grained simulations, the filter width can vary considerably. In fact,  filter sizes  define these methods. For example, in case of a LES of channel flow that is performed on a structured grid, the filter size in the span-wise and stream-wise directions scale with the wall-units and can be a constant. However, the filter width in the wall-normal direction can vary from a few wall units near the wall to $0.1\delta$ at the center of the channel or the edge of the boundary layer. In the traditional WMLES, the filter width is approximately of the order of $0.1\delta$ throughout in all directions. For hybrid RANS-LES  approaches (such as DES and IDDES), the filter width is of the order of $\delta$ in the span-wise and stream-wise directions, and similar to LES in the wall-normal direction. The non-uniform filtering requirement in LES and hybrid RANS-LES (HRLES) methods stems from the fact that in both the cases, the wall-stress is resolved which requires a near-wall grid that scales with wall units. In addition to  the filter size, the type of filter can also change the nature of the solution. For example, both the box and spectrally (sharp) filtered DNS both qualify as synthetic LES solutions.  In the case of finite element projections, the quality of the filter is linked to the order of polynomial used to filter the solution. In this work, we aim to address some of these issues by using finite element projections which allow for variation in the filter width in the domain and also provide the required flexibility to change the quality of the filter by changing the order of the polynomial. 

The idea of projection is at the core of the variational multiscale method (VMS) \citep{hughes1998variational,OSS,OSS2,VMS3,VMSE,NLVMS,NLVMS2,MZVMS}. In VMS,  projections are used to formally distinguish the coarse-scales from the fine-scales. The coarse-scale (filtered) solution that is obtained after the projection operation represents the `best' coarse-grained solution $u$ on the coarse-space based on some optimality condition, for example, the $L_2-$optimality condition. {\color{black} The sharp spectral filter obtained by truncation in Fourier space is also based on the idea of $L_2$-projection on to the Fourier basis functions}. In this work, we perform $L_2$-projections on finite element basis functions. It is also pertinent to mention that the current idea of optimal projections should not be confused with optimal LES \citep{optimal}. Our work optimally represents the DNS solution $u$ on a finite dimensional coarse-space, whereas,  optimal LES is an ideal LES model that targets accurate single-time multi-point statistics of the coarse solution.  

{\color{black} Projected DNS data have been previously used to improve both existing finite element methods \citep{pradhan2021variational} and turbulence models \citep{vreman1995priori,bou2008scale,meneveau2000scale,girimaji2005partially}. An important question is whether the applicability of the present filtering method is  only restricted to the finite element method because the functions that are used for projection are the finite element basis functions. In this paper, however, we employ them as an alternative to the traditional filters for assessing all kinds of methods and not just finite element methods. As discussed previously, the present approach has advantages in cases where the filter length is anisotropic, varies rapidly, or when non-homogeneous directions are present, as in wall-bounded flows.}
The filtering strategy that has some similarities to the present approach is the differential filter \citep{germano1986differential,najafi2015high}, which consists of a filtering length scale $l_p$. This length scale $l_p$ can be varied along the domain to have a similar effect.

In the past few years, several efforts have been made to train sub-grid models  using machine learning approaches both in an offline and model-consistent setting \citep{maulik2017neural, maulik2018data,maulik2019sub,beck2019deep, sarghini2003neural,gamahara2017searching,wang2018investigations,xie2019modeling,xie2019artificial,xie2020modeling}. Such data-driven LES models require a filtered form of the DNS, which in turn will depend on the filter size. Similarly, in WMLES, the model cannot be trained using the  mean solution in the first few grid points where the influence of the slip condition will be observed.  By applying projections and obtaining statistics from the optimal solution, one can obtain more reasonable targets to training the model \citep{beck2019deep,duraisamy2019turbulence,chung2022optimization} on more complex problems. Note that this is a first step towards addressing model consistency~\cite{duraisamy2021perspectives}.  Finite element projections are not restricted to simple geometries (Appendix B) and can be applied to more complex flows, and  provide the additional flexibility of choosing  polynomial orders for the geometry and the solution independently.

The main objective of the present work is to develop a unified framework that can be used as a lens to quantitatively assess, augment
and calibrate a wide range of coarse-grained models. Particular attention is paid to the behavior of various models in the proximity of the wall, and to ascertain whether scaling relationships exist. 

In section 2 of this paper, we describe the procedure of performing $L_2$-projection and provide a discussion on the choice of coarse basis functions that will be used. In section 3, we compute filtered solutions for the channel flow problem in the LES, WMLES and hybrid RANS-LES limit, and compute the coarse-scale statistics in each of these cases. In section 4, we show that the slip velocity in case of WMLES is a natural consequence of under-resolution in the wall-normal directions and guiding principles for improved slip wall models are proposed.  In section 5, we propose new slip-wall based wall model forms and evaluate its performance in comparison to traditional WMLES. Perspectives on improved slip-wall models are provided in section 6. Finally, we conclude the work in section 7.}

\section{Finite element projection}\label{sec:rules_submission}
\begin{figure}
    \centering
    \includegraphics[width=0.6\textwidth]{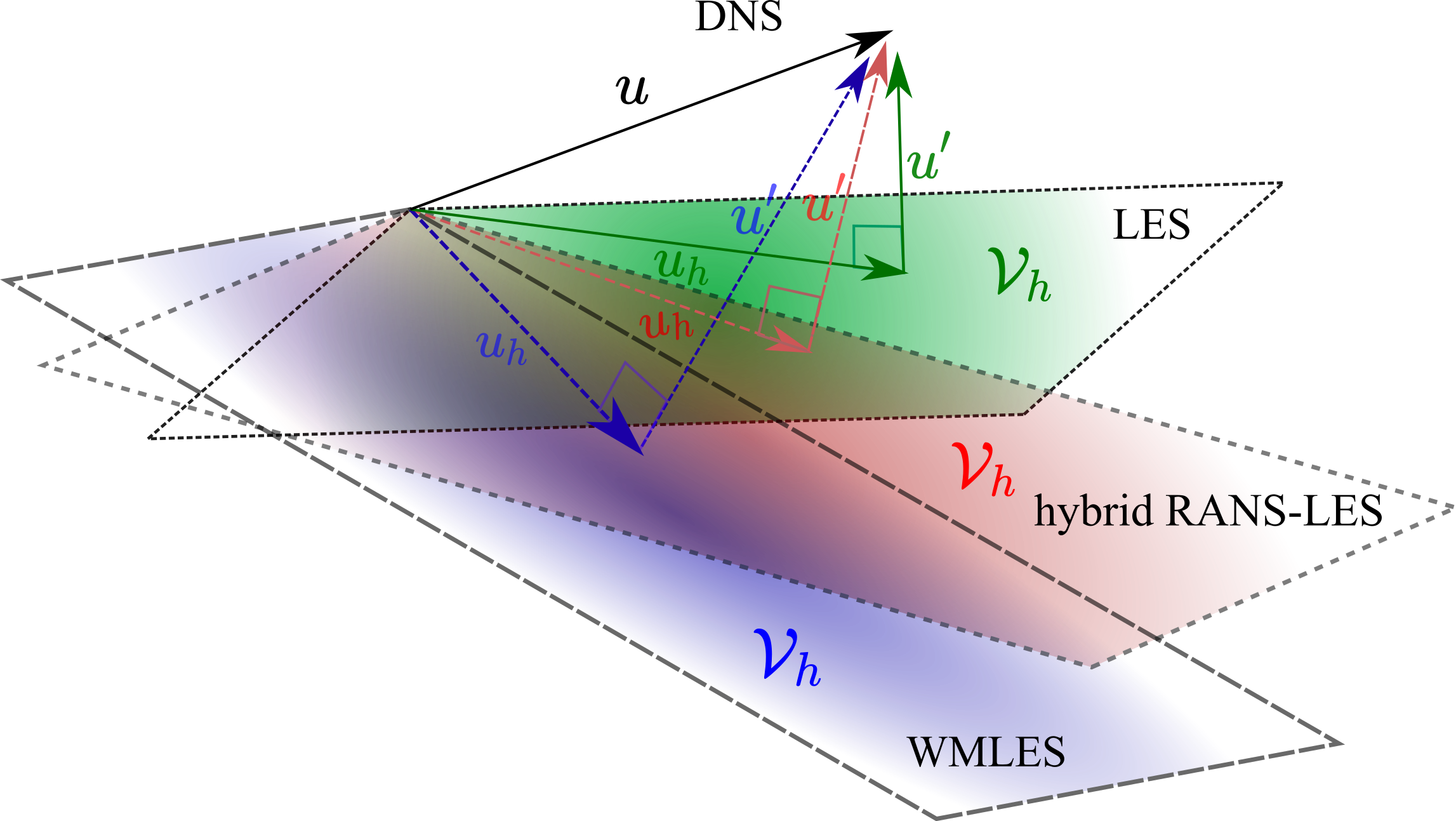}
    \caption{Schematic of the projection of the DNS solution $u$ on coarse finite element spaces $\mathcal{V}_h$ to obtain the $L_2-$optimal LES, WMLES or hybrid-RANS solution $u_h$.}
    \label{fig1PROJ}
\end{figure}
{\color{black} The goal of this section is to construct a generalised filtering approach that can be used to assess various coarse-grained simulations in problems using localized bases applicable to non-periodic boundary conditions. Further,  the meshes can contain an-isotropic elements along  with the possibility of grid-stretching. As a first step, however, we obtain high-resolution data for filtering.} For our purpose, we use the channel flow DNS data at friction Reynolds numbers of $Re_{\tau}\approx 1000$ and $Re_{\tau}\approx 5200$ from the Johns Hopkins Turbulence Database (JHTDB) \citep{li2008public,lee2015direct} and a smaller channel $Re_{\tau}\approx 950$ case form the Texas turbulence file server \citep{hoyas2006scaling}. To this end, consider the decomposition of the full-order (DNS) solution  $u$ into coarse and fine scales as 
\begin{equation}
u = {u_h} + u',
\end{equation}

where ${u_h} \in {\mathcal{V}_h}$ and $u' \in \mathcal{V}'$ as shown in Fig. \ref{fig1PROJ}. The vector space of functions $\mathcal{V}\equiv {L}^2(\Omega)$ is the space of  square-integrable functions. This space is  decomposed as 
\begin{equation}
\mathcal{V} = {\mathcal{V}_h} \oplus \mathcal{V}',
\end{equation}
where $\oplus$ represents a direct sum of ${\mathcal{V}_h}$ and $\mathcal{V}'$. Let us also define $\mathcal{T}_{h}$ to be a tessellation of domain $\Omega$ into a set of non-overlapping elements, $K$, each having a sub-domain $\Omega_k$ and boundary $\Gamma_k$. The functional space $\mathcal{V}$ is infinite dimensional and must be approximated by a finite dimensional approximation ${\mathcal{V}_h}$. The domain and boundary of an element marked by $\Omega_e$ and $\Gamma_e$ respectively.  In case of the continuous Galerkin (CG) method, the coarse space basis functions ${\mathcal{V}_h} \subset C^{0} \cap {L}^2(\Omega)$ have $C^{0}$ continuity everywhere including element boundaries. In the case of  discontinuous Galerkin (DG) methods, the coarse space ${\mathcal{V}_h}$ is defined as
\begin{equation}
{\mathcal{V}_h} \triangleq \left\{u\in{L_2}(\Omega):u|_{T}\in P^k(T),T\in\mathcal{T}_{h} \right\},
\end{equation}
where the space of polynomials up to degree $k$ is denoted as $P^k$. Defining ${\mathcal{V}_h}$ in this manner allows for discontinuities in the solution across element boundaries. The DG space is a more richer space compared to CG space if both the number of elements and polynomial order are kept fixed. Irrespective of the choice of basis functions used (CG or DG), given $u$ from the high-fidelity simulation, our goal is to find the optimal representation of $u$ in the coarse sub-space ${\mathcal{V}_h}$. In our case, we will use the ${L}^2$-projection to obtain $u_h$ which minimises the value of $||u-u_h||_2^2$. This problem is equivalent to the problem of finding $u_h\in {\mathcal{V}_h}$ such that 
\begin{equation}
(u,w_h)=(u_h,w_h)  \quad \forall {w_h} \in {\mathcal{V}_h}.
\end{equation}
{\color{black} where $(\cdot,\cdot)$ denotes the $L_2$ inner product and $w_h$ denotes a coarse-space weighting function}. In the case of CG basis functions, the mass matrix  is global and a large matrix needs to be inverted to obtain the final filtered solution. The DG mass matrix on the other hand is local to the element and lends itself to easy parallelization. 

 Since $u_h$ and $w_h$ are finite dimensional, their inner product can be computed precisely using quadrature rules. $(u,w_h)$ requires  special care because $u$ is extremely high dimensional in comparison to $w_h$. The high-dimensionality of $u$ is restricted by the size of DNS which exists on a very fine mesh capable of resolving the Kolmogorov scales O($\eta$). To compute this term precisely, we interpolate the coarse-scale basis functions and the DNS solution on a very fine mesh of the size of O($\eta$) and apply numerical integration to compute the inner products. The size of the numerical integration mesh is adjusted till the final projected solution is independent of the numerical integration mesh size. Additional details on the procedure to compute the $L_2$-procedure is given in Appendix \ref{appendixA}.

The final comment is on the imposition of the near-wall behavior of the coarse-space. There are two choices: (i.)  project on a space which strongly satisfies the boundary condition at the nodal points, or (ii.) keep the boundary DOFs free and make no such assumptions. The second choice appears more reasonable because when the solution is coarse-grained in the wall-normal direction, the solution might no longer satisfy the boundary conditions strongly. This is especially true for WMLES where the coarse-solution no longer satisfies the boundary condition and slip is observed. However, as the grid is refined near the wall, the no-slip boundary condition is naturally satisfied.  

\section{Application to channel flow.}\label{sec:types_paper}

As a first step towards obtaining the projected DNS solution for the channel flow problem, we discuss the effect of the choice of the coarse-space basis functions on the coarse-scale solution obtained after the projection operation. Depending on the coarse-space basis, the projected solution can be either a low-dimensional compressed representation of the original solution or a spatially filtered version of it. The low-dimensional compressed representation is obtained when the coarse basis is tailored using data or existing analytical solutions. To ensure that the projection step leads to a more general spatial filtering approach, non-tailored basis functions commonly used in the finite element method are used. The projection operation onto these coarse finite element grids will lead to filtering.

The resulting coarse solution after filtering might be considerably different from the  DNS solution due to truncation of the high-frequency components present originally in the DNS solution. In the near-wall region, the effect of projection can vary with the size of the filter in each direction. One manifestation of under-resolution in the wall-normal direction is the occurrence of a slip velocity. This slip-velocity can - in fact - be tracked down to the mean-profile itself. To resolve the mean solution, a near-wall grid spacing of $\Delta y^+ \approx 1$ is required in the wall-normal direction. However, if a grid size of $\Delta y \approx 0.1\delta$ is used, even the mean solution can no longer be resolved and a slip velocity at the wall will be observed. This is true unless the solution is artificially forced to go from a large value to zero over just one grid point. A solution to make the coarse-scale solution satisfy the no-slip boundary condition is to enrich the coarse-space with a tailored basis \citep{krank2016new}. As a consequence, the tailored basis mimics the mean profile between the wall and the first grid point and ensures that the no-slip is satisfied collectively by the coarse non-tailored basis and the enriched tailored basis. 

To define the coarse space, we first construct a finite element mesh and chose the polynomial order of the basis functions. The idea here is that by selecting the grid and the polynomial order of the basis functions, we are enforcing our desired filter size distribution. A variety of coarse spaces have been generated as shown in table \ref{tab:kd}. The 'A'-type grids are the DNS grids on which the high-resolution solution $u$ exists. Two different DNS solutions at friction Reynolds numbers of $Re_\tau \approx 950$ and  $Re_\tau \approx 1000$ are used and their corresponsing grids are marked as $A1$ and $A2$, respectively. The  $Re_\tau \approx 950$ solution \citep{hoyas2006scaling,del2004scaling} is obtained from a relatively smaller domain having a stream-wise size of $L_x \approx 2\pi\delta$ and a span-wise sizes of $L_z \approx \pi\delta$ , whereas, the $Re_\tau \approx 1000$ solution \citep{li2008public,perlman2007data,lee2015direct} is obtained as a cutout from a simulation performed on a larger domain. 
 
  The 'B'-type grids, on the other hand, are tailored for performing wall-resolved LES. As a result of the size of the largest energy containing eddies scaling with the distance from the wall \citep{yang2021grid} outside the viscous sub-layer, a mesh resolution of $\Delta y \approx 0.1\delta-0.25 \delta$ is used for the type-'B' grids at the center of the channel. Similarly, the 'C'-type grids are tailored for performing WMLES simulations using the wall-stress or the slip-wall based approaches.  The type-'D' grids are more suitable for assessing the WMLES branch of the hybrid RANS-LES methods \citep{shur2008hybrid}. For type-'D', the resolution in the stream-wise and the span-wise direction is similar to the 'C'-type grid. However, in the wall-normal direction, a grid spacing similar to type-'B' grid has been assumed i.e. $\Delta y^+ \approx 0.1-1$  in the near wall region and $\Delta y \approx 0.1\delta-0.25 \delta$ at the center of the channel. The type-'E' grid is an extremely coarse grid with resolutions of $\Delta x \approx 0.35\delta$,$\Delta y \approx 0.334\delta$ and $\Delta z \approx 0.35\delta$ in the stream-wise, span-wise and the wall-normal directions, respectively. For all the grid types, the mesh is uniform in the stream-wise and the span-wise directions. However, in the wall normal direction, the mesh has been stretched geometrically for cases 'B' and 'D'. In case of 'C' and 'E' type grids, uniform mesh is assumed in the wall normal directions as well. For each type of grid, two different polynomial orders $p=1,2$ are used to construct the projection coarse-space. The stretch rates (SR) and the polynomial orders for different cases have been summarised in table \ref{tab:kd}. 

\begin{figure}
    \centering
    \includegraphics[width=1.00\textwidth]{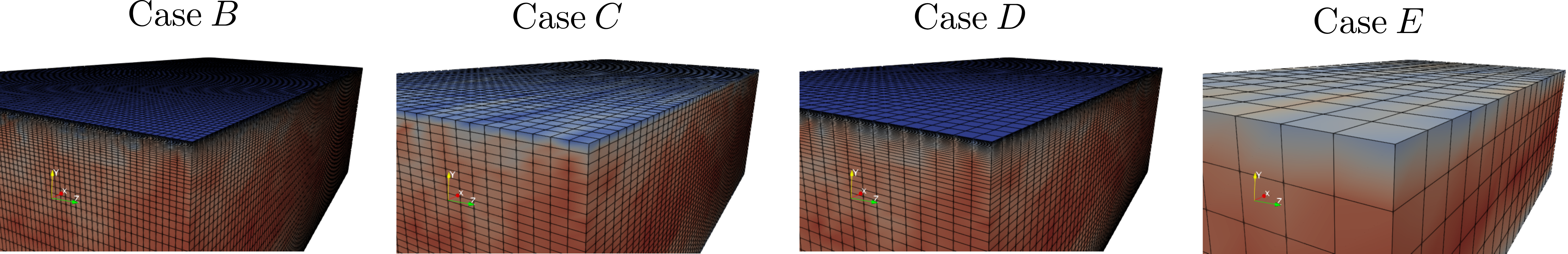}
    \caption{Near-wall grids used for cases B,C,D and E.}
    \label{figgridBCDE}
\end{figure}

\begin{figure}
	
\centering
\includegraphics[width=1.0\textwidth]{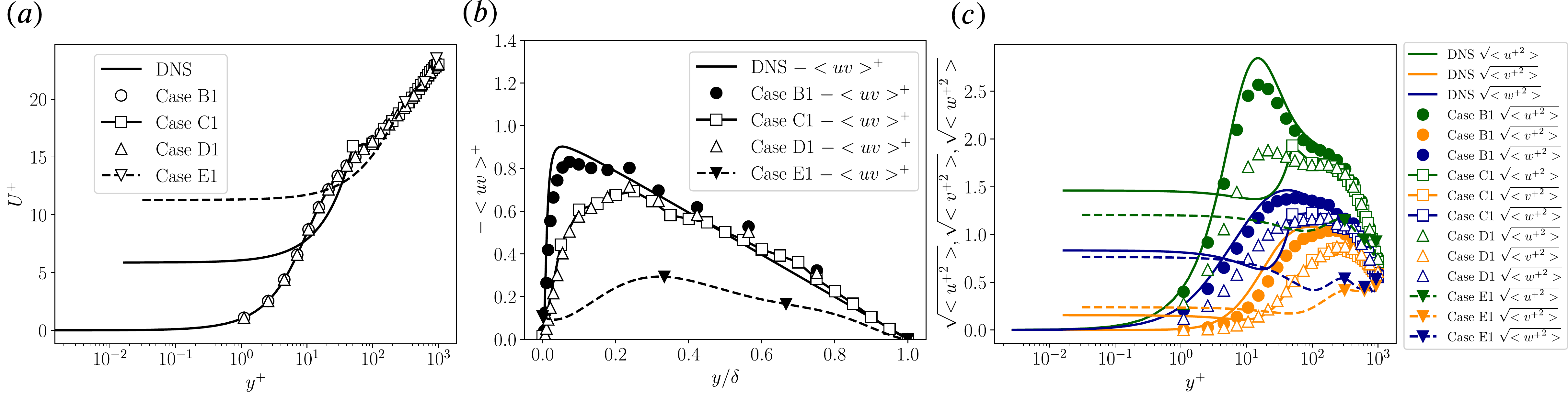}
\caption{Comparison of mean and second-order statistics for cases B1,C1,D1 and E1. The symbols in all the plots correspond to the value at the nodal point. In sub-plot (a),  the solutions for cases C1 (WMLES) and E1 (extremely coarse), are interpolated to the DNS mesh using the coarse finite element basis functions near the wall to show the slip effects.}
\label{fig1mix}
\end{figure}

Fig. \ref{fig1mix} shows the  mean and second-order statistics computed using the projected solution for the cases B1,C1,D1 and E1. The goal is to compare the optimal solutions for different grids that correspond to different coarse-grained approaches, except for E1, which is an extremely coarse mesh and is not suitable for any existing method. From Fig. \ref{fig1mix}(a), it can be observed that the mean velocity is well-resolved for cases B1 and D1. For case C1, which represents an optimal WMLES solution, the mean velocity is well-resolved only after the first grid point i.e. $y/\delta > 0.05$. As can be observed in \ref{fig1mix}(a), Case E1 is extremely coarse and fails to resolve the mean velocity until the outer limit of the log-layer is reached. For these cases with wall-normal under-resolution, the effect of under-resolution results in slip velocity $u_s$ at the wall, {\color{black} which can be calculated by evaluating $u_h$ at the wall}. The magnitude of the mean stream-wise slip velocity $<u_s>^+$ was found to increase with the under-resolution i.e. $<u_s>^+ \approx 6$ for case C1 to $<u_s>^+ \approx 12$ for case E1. It can also be observed that all the methods except E1  resolve the second-order statistics outside the inner layer. Inside the near-wall region, only B1 is capable of accurately resolving the turbulence stresses. Among the second-order statistics, the effect of filtering is most strongly felt on the wall-normal fluctuations. It can be observed for  cases C1 and D1 that the wall-normal fluctuations far away from the inner layer are under-represented even when  stream-wise and the span-wise fluctuations as close to the DNS solution.  

\begin{figure}
    \centering
    \includegraphics[width=1.0\textwidth]{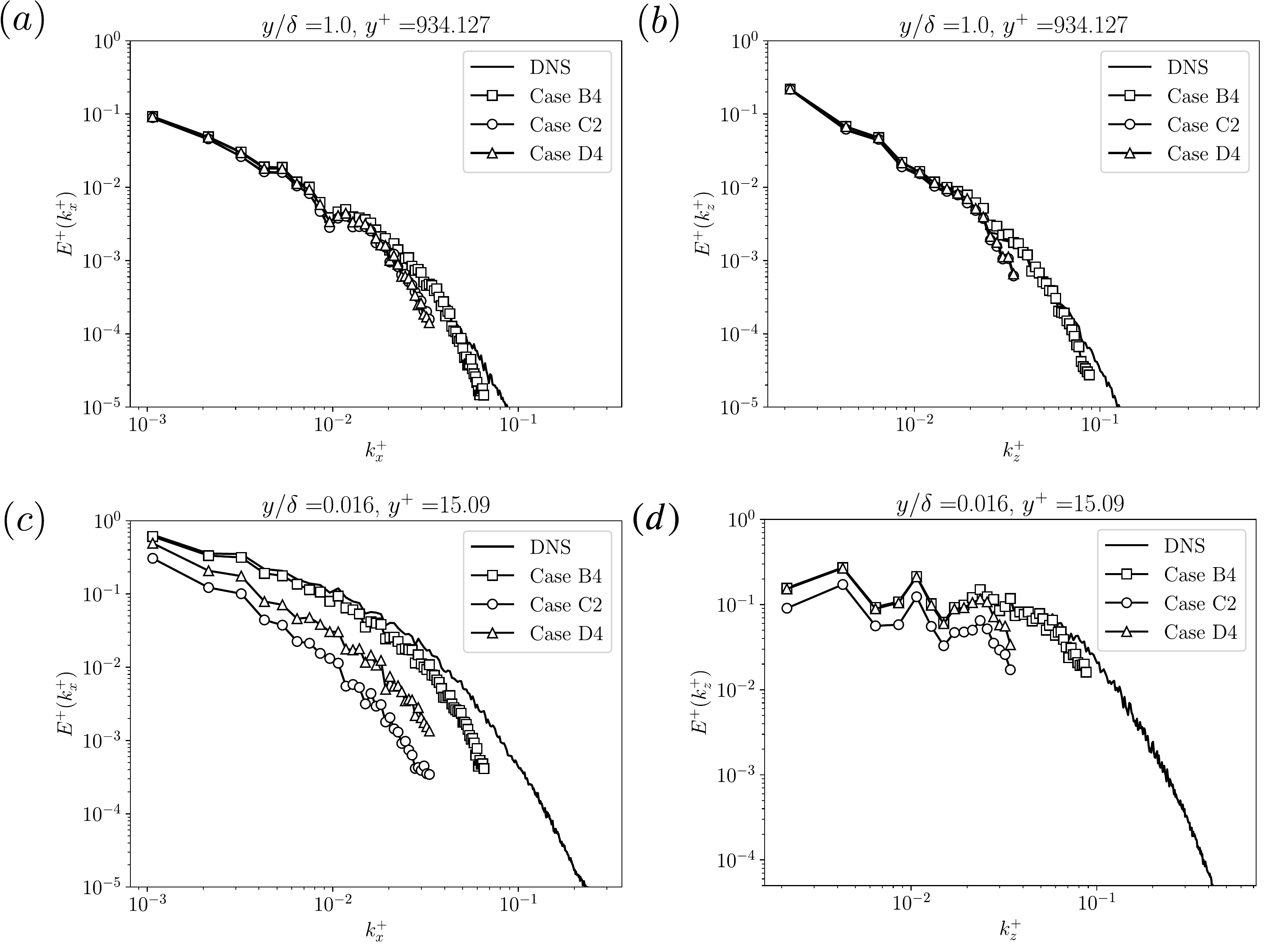}
    \caption{Comparison of the energy spectra for the stream-wise velocity component at the near-wall region and the center of channel for cases A2,B4,C2 and D4.}
    \label{fig1SPECTRA}
\end{figure}

{\color{black} Figure \ref{fig1SPECTRA} shows the stream-wise velocity energy spectra of the projected solution in the span-wise and stream-wise directions for cases A2, B4, C2 and D4 at two different wall-normal locations. The choice of cases plotted here is based on the most suitable mesh sizes for the various methods. As can be seen from figures \ref{fig1SPECTRA}(a) and \ref{fig1SPECTRA}(b), the large scales are well-represented at the center of the channel ($\frac{y}{\delta} \approx 1.0$) both in the stream-wise and the span-wise directions, for all the methods.  However, as can be seen in figure \ref{fig1SPECTRA}(c), the large scales are not represented accurately by the C2 (WMLES) and D4 (hybrid RANS-LES) cases in the near-wall region ($y^+ \approx 15$). However, in the span-wise direction (figure \ref{fig1SPECTRA}(d)), the large scales are relatively well represented in D4 in comparison to C2. In the next part, the effect of the projection will be discussed individually for each type of mesh.}

%\begin{figure}
%    \centering
%    \includegraphics[width=1.0\textwidth]{./plot.pdf}
%    \caption{Comparison of mean and second-order statistics for cases B1,C1,D1 %and E1. The symbols in all the plots correspond to the value at the nodal %point. In sub-plot (b),  the solutions for cases C1 (WMLES) and E1 (extremely %coarse), are interpolated to the DNS mesh using the coarse finite element basis %functions to show the slip velocity. Identical symbols are used for the %sub-plots (c) and (e).}
%    \label{fig1mix}
%\end{figure}

%\begin{figure}
%	
%	\begin{subfigure}[b]{0.3\textwidth}
%		\centering
%		\includegraphics[width=\textwidth]{./compare_method/2.pdf}
%	\end{subfigure}
%    %
%    \begin{subfigure}[b]{0.3\textwidth}
%    	\centering
%    	\includegraphics[width=\textwidth]{./compare_method/4.pdf}
%    \end{subfigure}
%    %
%    \begin{subfigure}[b]{0.45\textwidth}
%    	\centering
%    	\includegraphics[width=\textwidth]{./compare_method/5.pdf}
%    \end{subfigure}
%\caption{Comparison of mean and second-order statistics for cases B1,C1,D1 and E1. The symbols %in all the plots correspond to the value at the nodal point. In sub-plot (a),  the solutions %for cases C1 (WMLES) and E1 (extremely coarse), are interpolated to the DNS mesh using the %coarse finite element basis functions to show the slip velocity.}
%\label{fig1mix}
%\end{figure}

\begin{table}
  \begin{center}
\def~{\hphantom{0}}
  \begin{tabular}{lccccccccccccc} 
     Case & $Re_\tau$ & $N_x \times N_y \times N_z$ & $p$ & $SR$ & $\Delta_x^+$ & ${\Delta_x}/{\delta}$  &  $\Delta_y^+$ & ${\Delta_y}/{\delta}$  & $\Delta_z^+$ & ${\Delta_z}/{\delta}$ \\[3pt]
     A1 & 1000 & $512 \times 512 \times 512$ & Sp. & -    & 12.24 & 0.0122 & 0.016 $-$ 6.14   & 1.65E-5 $-$ 0.006 & 6.12  & 0.006\\
A2 & 950  & $512 \times 384 \times 512$ & Sp. & -    & 11.47 & 0.0122 & 0.031 $-$ 7.64   & 3.35E-5 $-$ 0.008 & 5.73  & 0.006\\
B1 & 1000 & $121 \times 41  \times  81$ & 1   & 1.33 & 39.11 & 0.0391 & 1.10  $-$ 248.45 & 0.001   $-$ 0.248 & 39.11 & 0.039\\
B2 & 950  & $121 \times 41  \times  81$ & 1   & 1.33 & 36.71 & 0.0393 & 1.03  $-$ 232.55 & 0.001   $-$ 0.248 & 36.71 & 0.039\\
B3 & 950  & $121 \times 81  \times  81$ & 2   & 1.33 & 36.71 & 0.0393 & 0.51  $-$ 116.27 & 5.54E-4 $-$ 0.124 & 36.71 & 0.039\\
B4 & 950  & $121 \times 97  \times  81$ & 1   & 1.10 & 36.71 & 0.0393 & 0.97  $-$ 85.80  & 0.001   $-$ 0.091 & 36.71 & 0.039\\
C1 & 1000 & $61  \times 41  \times  31$ & 1   & 1.00 & 104.3 & 0.1045 & 49.90            & 0.05              & 104.3 & 0.1045\\
C2 & 950  & $61  \times 41  \times  31$ & 1   & 1.00 & 97.91 & 0.1048 & 46.7             & 0.05              & 97.91 & 0.1048\\
C3 & 950  & $61  \times 81  \times  31$ & 2   & 1.00 & 97.91 & 0.1048 & 23.3             & 0.025             & 97.91 & 0.1048\\
C4 & 950  & $61  \times 97  \times  31$ & 1   & 1.00 & 97.91 & 0.1048 & 19.46            & 0.021             & 97.91 & 0.1048\\
D1 & 1000 & $61  \times 41  \times  31$ & 1   & 1.33 & 39.11 & 0.1045 & 1.10  $-$ 248.45 & 0.001   $-$ 0.248 & 104.3 & 0.1045\\
D2 & 950  & $61  \times 41  \times  31$ & 1   & 1.33 & 97.91 & 0.1048 & 1.03  $-$ 232.55 & 0.001   $-$ 0.248 & 97.91 & 0.1048\\
D3 & 950  & $61  \times 81  \times  31$ & 2   & 1.33 & 97.91 & 0.1048 & 0.51  $-$ 116.27 & 5.54E-4 $-$ 0.124 & 97.91 & 0.1048\\
D4 & 950  & $61  \times 97  \times  31$ & 1   & 1.10 & 97.91 & 0.1048 & 0.97  $-$ 85.80  & 0.001   $-$ 0.091 & 97.91 & 0.1048\\
E1 & 1000 & $19  \times 7   \times  10$ & 3   & 1.00 & 326.3 & 0.3500 & 311.27           & 0.334              & 326.3 & 0.35\\
  \end{tabular}
  \caption{Summary of mesh parameters. Here, $\Delta_x^+$,  $\Delta_y^+$ and $\Delta_z^+$ are the effective grid sizes in different directions $\Delta_x$, $\Delta_y$ and $\Delta_z$ normalised with wall units, $\delta$ is the half channel height, $N_x$, $N_x$ and $N_z$ represents the number of degrees of freedom in the stream-wise, wall-normal and span-wise directions respectively, $p$ is order of polynomial used, $SR$ is the stretching ratio used to generate the grid. The effective grid sizes $\Delta_x$, $\Delta_y$ and $\Delta_z$ for the finite element grid are defined as $\Delta_x = \Delta^e_x/p$, $\Delta^e_y/p$ and $\Delta^e_z/p$ respectively. The quantities $\Delta^e_x$, $\Delta^e_y$ and $\Delta^e_z$ represent the actual element sizes in the finite element mesh.}
  \label{tab:kd}
  \end{center}
\end{table}

\begin{figure}
\includegraphics[width=1.00\textwidth]{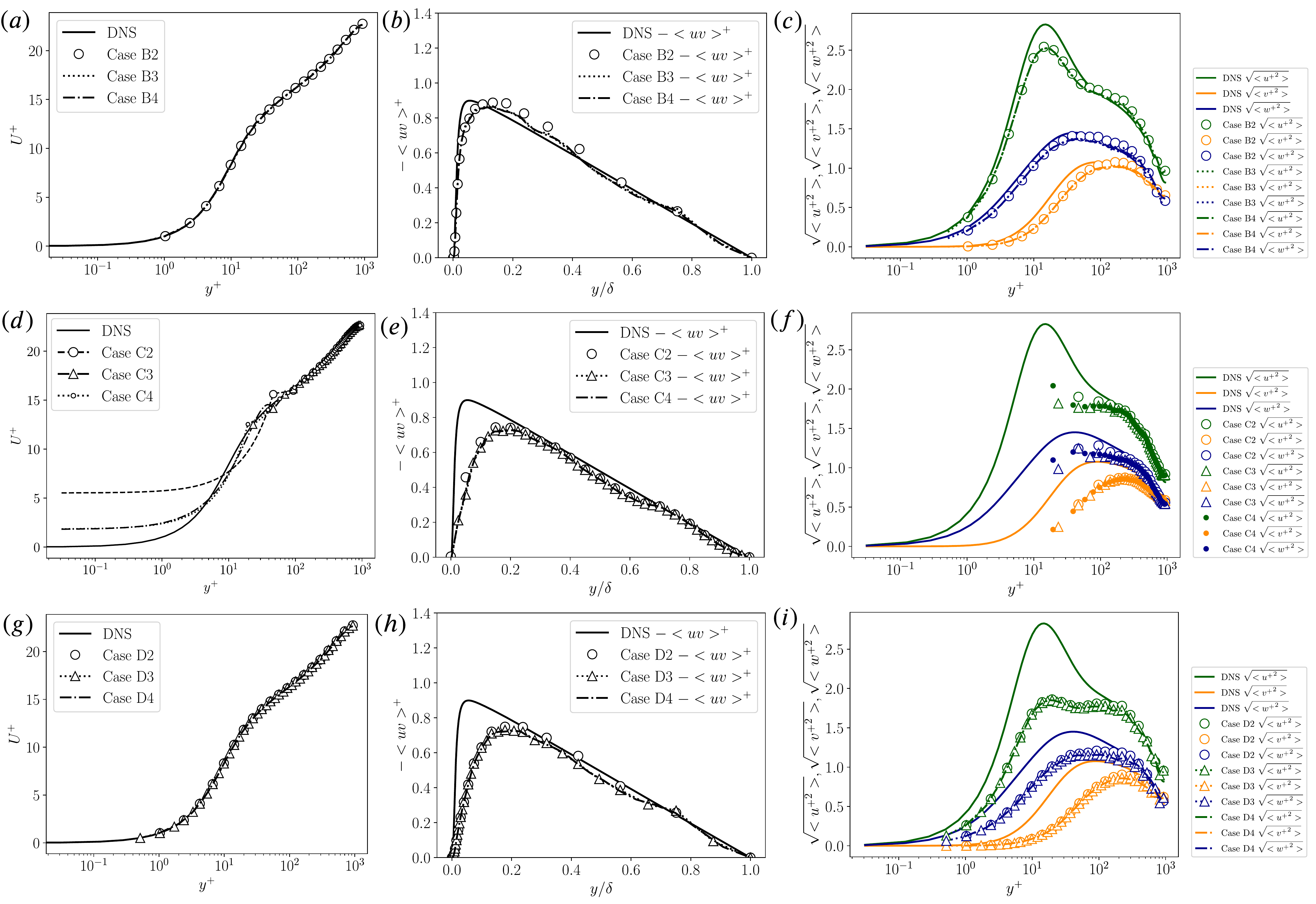}
\caption{Comparison of optimal projections. Upper row: Wall-resolved cases; Middle row:  WMLES; Bottom row: HRLES.}
\label{fig1LES}
\end{figure}

\noindent {\bf The wall-resolved LES limit: }  Coarse-scale statistics for B2, B3 and B4 are provided in the top row of Fig. \ref{fig1LES}. It can be observed  that all the cases perform well in resolving the mean profile. Similar trends are  observed in Figs. \ref{fig1LES}(b), \ref{fig1LES}(c), where the cases B3 and B4 only slightly outperform the case B2 in resolving the second-order statistics. This suggests that the sensitivity of the coarse-scale statistics to the wall-normal stretch rate is not as high as the sensitivity to the mesh resolution in the stream-wise and the span-wise directions.   

%\begin{figure}
%	\begin{subfigure}[b]{0.3\textwidth}
%		\centering
%		\includegraphics[width=\textwidth]{./LES_method/1.pdf}
%	\end{subfigure}
%	%
%	\begin{subfigure}[b]{0.3\textwidth}
%		\centering
%		\includegraphics[width=\textwidth]{./LES_method/2.pdf}
%	\end{subfigure}
%    %
%    \begin{subfigure}[b]{0.4\textwidth}
%    	\centering
%    	\includegraphics[width=\textwidth]{./LES_method/3.pdf}
%    \end{subfigure}
%    
%      \begin{subfigure}[b]{0.3\textwidth}
%		\centering
%		\includegraphics[width=\textwidth]{./WMLES_method/2.pdf}
%	\end{subfigure}
%		\begin{subfigure}[b]{0.3\textwidth}
%		\centering
%		\includegraphics[width=\textwidth]{./WMLES_method/4.pdf}
%	\end{subfigure}
    %
%    \begin{subfigure}[b]{0.3\textwidth}
%		\centering
%		\includegraphics[width=\textwidth]{./WMLES_method/3.pdf}
%	\end{subfigure}
    %

%	 \begin{subfigure}[b]{0.3\textwidth}
%		\centering
%		\includegraphics[width=\textwidth]{./HRLES_method/2.pdf}
%	\end{subfigure}
%	\begin{subfigure}[b]{0.3\textwidth}
%		\centering
%		\includegraphics[width=\textwidth]{./HRLES_method/4.pdf}
%	\end{subfigure}
    %
%    \begin{subfigure}[b]{0.4\textwidth}
%    	\centering
%    	\includegraphics[width=\textwidth]{./HRLES_method/5.pdf}
%    \end{subfigure}
    
%\caption{Comparison of optimal projections. Upper row: Wall-resolved cases; Middle row:  WMLES; %Bottom row: HRLES.}
%\label{fig1LES}
%\end{figure}

\noindent {\bf The WMLES limit: }
Coarse-scale statistics for C2, C3 and C4 are provided in the middle row of Fig. \ref{fig1LES}. In cases C2, C3 and C4, the mesh resolutions in wall units is much larger than unity and cannot resolve the mean velocity profile accurately near the wall. As a consequence, a slip velocity $u_s$ is observed for all the cases after the projection step. It was also observed that the magnitude of the mean slip velocity $<u_s>^+$ is highest for the case with the maximum wall-normal under-resolution i.e. C2. This magnitude goes down as the wall-normal mesh is refined from C2 to C3 or C2 to C4. As expected, the second-order coarse-scale statistics are only accurate outside the inner layer towards the center of the channel. It can also be observed that the stream-wise and span-wise velocity fluctuations computed using the coarse solution do not go to zero near the wall. The wall-normal velocity fluctuations, however, go to zero at the wall. This near-wall behavior of the coarse-scales is consistent with  existing WMLES simulations in the literature e.g. \cite{wang2020comparative,kawai2012wall}.

\noindent {\bf The Hybrid RANS-LES limit: }
Cases D2, D3 and D4  represent the grids for hybrid RANS-LES (HRLES) methods such as IDDES \citep{shur2008hybrid}. Coarse-scale statistics for D2, D3 and D4 are provided in the bottom row of Fig. \ref{fig1LES}. The mean profile is resolved accurately. The second-order statistics in the region outside of $10-20\%$ of the boundary layer is almost identical to that of the 'C'-type grids in the middle row. However, near the wall, unlike the 'C'-type WMLES cases, all the velocity fluctuations go to zero due to the no-slip condition being satisfied by the hybrid RANS-LES cases, and they are under-represented when compared to 'B'-type LES cases. These observations are consistent with previous results from the literature \citep{friess2020formulation}.  

{\color{black} In this section, we have used the labels LES, WMLES and HRLES above to distinguish between  the various methods. However, the difference between the later two is subtle. The traditional WMLES method essentially uses RANS knowledge to compute the wall-stress at the wall and can  also be called a hybrid RANS-LES approach. Similarly, the HRLES approaches (such as IDDES) by virtue of solving the RANS equations near the wall reduce the computational cost associated with resolving the  wall and can  also be considered a WMLES. However, the context in which the labels WMLES and HRLES have been used in this paper is based on whether these models integrate to the wall or not in a single domain, in other words if the size of the filter in the wall-normal direction is large or not in wall-units. } 

\begin{figure}
\includegraphics[width=1.00\textwidth]{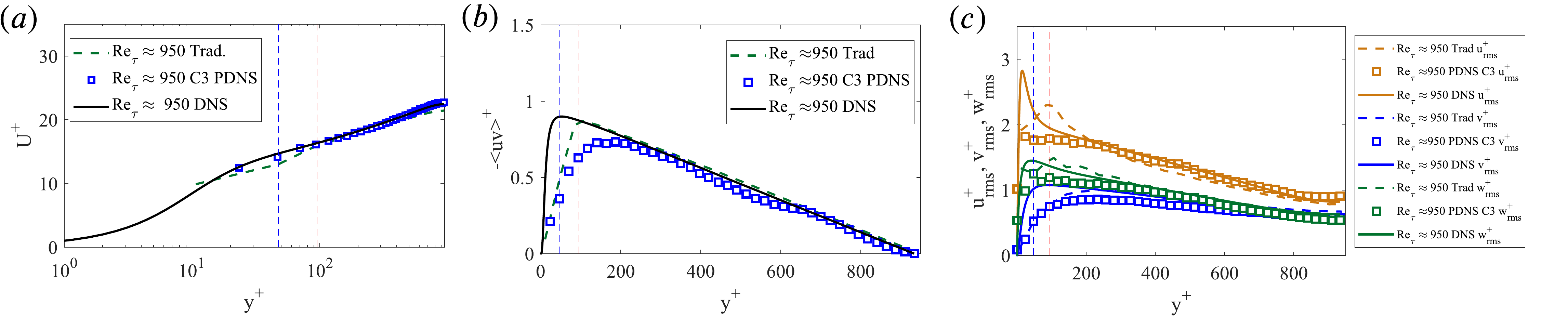}
\caption{Comparison of projected DNS and solution from the traditional WMLES method at similar resolutions.}
\label{fig6LES}
\end{figure}

{\color{black}
It is worthwhile to mention that the results presented in this section appear to be more accurate in comparison to those in the literature. For instance, a large stretching ratio of 1.33 has been used for some of the meshes which do not induce a significant error in the filtered solution, however, this stretch rate is more than the suggested limit for many methods. Additionally, no log-layer mismatch (LLM) was obtained in any of these cases. The resolution considered here for WMLES and hybrid RANS/LES is of the order of 10 points per semi-channel height $\delta$ in the streamwise and the spanwise directions, which is coarse compared with the guidelines for these approaches. A comparison between the optimal solution for the C3 case and a WMLES solution using the traditional wall-stress-based approach computed on the same grid is presented in figure \ref{fig6LES}. The mean solution for both cases begins to deviate from the DNS at similar locations. The resolved turbulent shear stress is under-represented near the wall, starting almost identically, however, differing in their peaks. Similarly, the wall-normal velocity fluctuations are almost identical. On the other hand, the velocity inside the first element is slightly under-predicted in the traditional WMLES approach in comparison to projected DNS. Additionally,  both the stream-wise and span-wise velocity fluctuations reveal an overshoot near the second off-wall grid point in the traditional WMLES method. 

One possible reason for the discrepancies between the results from the true simulation and the filtered DNS is the lack of accurate closures.  With a poor model, the LLM may persist until a DNS-like resolution is reached. This is - in principle - similar to attributing inaccuracies in a wall resolved LES with a standard Smagorinsky model when a dynamic Smagorinsky model might yield near-optimal performance. However, using the optimal projection framework presented here, it is now possible to perform an analysis of the closure terms and evaluate modeling errors. By reducing the modeling errors, the goal is to force the solution to reach a near-optimal state. Ideally, we would have wanted to improve all three approaches using our optimal projection framework. However, to have a compact presentation, we only consider evaluating the modeling errors in the slip-wall-based WMLES models and improve its a posteriori performance.   
}

\section{Analysis of slip-based wall models.}\label{sec:filetypes}
The slip-velocity at the wall in WMLES is related to  under-resolution in the wall-normal direction. In this section, we seek to quantify this slip-velocity to ensure that the resulting model generalises well to different Reynolds numbers. To understand the Reynolds number dependence, DNS from two different friction Reynolds numbers of $Re_{\tau} \approx 1000$ and $Re_{\tau} \approx 5200$ are used. As mentioned earlier, the computation of the 3-D projection of the $Re_{\tau} \approx 5200$ case by sequential 1-D projections in the wall-normal, stream-wise and the span-wise directions is computationally expensive. To ensure computational efficiency and  utilizing the fact that this is a near-wall phenomenon, we project the DNS solution on uniform elements of size $\Delta_e$ with polynomial basis functions. This is equivalent to performing a full 3-D projection on a DG finite element solution space. The element shares its bottom face with the wall of the channel to mimic a near-wall grid. By moving the position of this element on the wall surface, different realizations of the slip velocity and coarse solution gradients in the wall normal direction can be obtained. This is possible due to the statistical homogeneity present in the stream-wise and span-wise directions. 

\begin{figure}
     \centering
     \includegraphics[width=1.00\textwidth]{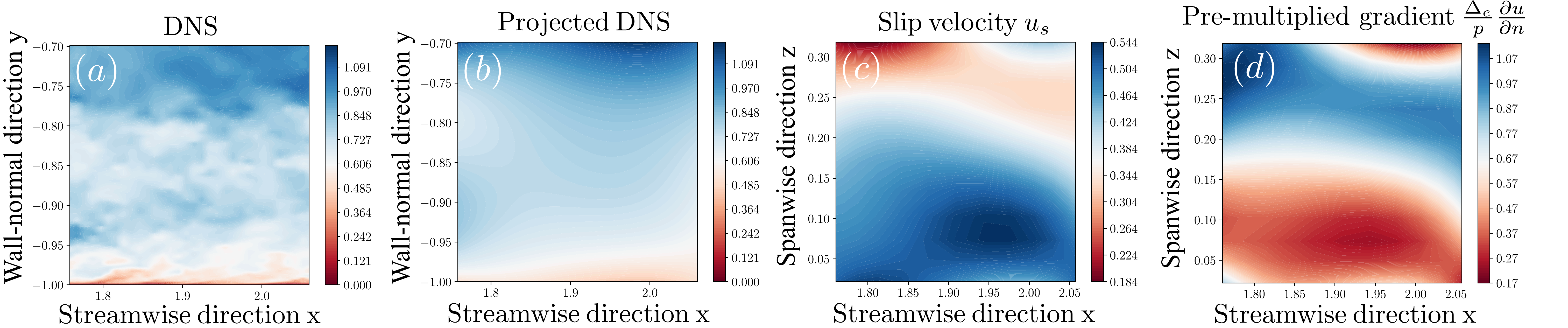}
     \caption{Filtered solution inside an sample element obtained by 3-D Projection of near-wall $Re_{\tau} \approx 5200$ channel data.}
     \label{DG_sample}
\end{figure}

For each realization, the projection of the DNS solution on the finite dimensional DG space leads to filtering of the DNS solution as shown in Fig. \ref{DG_sample}. Fig. \ref{DG_sample}(a) shows the contour of the DNS solution of the stream-wise velocity component for a sample 3-D element. The projected DNS solution for the same element is shown in Fig. \ref{DG_sample}(b). The projected DNS solution does not satisfy the no-slip boundary condition at the wall and does not contain the fine-scale information  present in the original DNS solution. The goal is to assess the slip-wall based wall model proposed by \cite{bose2014dynamic,bae2019dynamic} 
\begin{equation}
    u_s = C_w \Delta \frac{\partial u_h}{\partial n},
    \label{slipmod}
\end{equation}
and obtain an estimate of the model coefficient $C_w$. {\color{black} The coarse field $u_h$ can be obtained by either projecting the stream-wise, the span-wise or the wall-normal velocity fields. As a result, the value of $C_w$ obtained is tied to the velocity component that is used for projection}. The computation of $C_w$ using equation \eqref{slipmod} requires the computation of $u_s$ and the pre-multiplied gradient $\Delta \frac{\partial u_h}{\partial n}$. The slip-velocity $u_s$ is obtained by evaluating the coarse-scale solution at the wall as shown in  Fig. \ref{DG_sample}(c). The pre-multiplied gradient $\Delta \frac{\partial u_h}{\partial n}$ is obtained by computing the derivatives of the coarse-scale in the wall-normal direction and multiplying with the normalized resolution $\Delta = {\frac{\Delta_e}{p}}$ as shown in Fig. \ref{DG_sample}(d). However, this results in  an over-determined system for $C_w$. In general, the value of $C_w$ is also expected to change with the filter size $\Delta$ used for the projection operation. The problem of this system being  over-determined is solved by performing a least-squares minimization over many such realizations till convergence in the estimates of $C_w$ was obtained. To solve the problem of the model coefficient $C_w$ being dependent on the filter size $\Delta$, we perform dimensional analysis. Other parameters that could affect $C_w$ are a) the order of polynomial used for projection $p$, the viscosity $\nu$;  and b) the wall stress $\tau_w$. After non-dimensionalization, the following model form for $C_w$ can be obtained: $
    C_w = g_{p}(\Delta^+),$
where $g_{p}$ is a function of the grid resolution normalized with wall units $\Delta^+$ and the subscript $p$ denotes the coarse space polynomial order used for projection. The parameter $\Delta^+$ can be considered to be an indicator of the near-wall grid resolution. Similarly, the order of the numerical method can be encoded in $p$. Higher $p$ implies that a more accurate numerical method has been used to compute the LES solution. However, this implies that for every polynomial order $p$ we have to learn a new function. In addition to this, the numerical methods used to perform LES might work sub-optimally and the exact order might not be preserved. Hence, it is necessary that the effect of the numerical method be parameterized through a model constant similar to the Smagorinsky model coefficient $C_s$. 

\begin{figure}
	\begin{subfigure}[b]{0.5\textwidth}
		\centering
		\includegraphics[width=\textwidth]{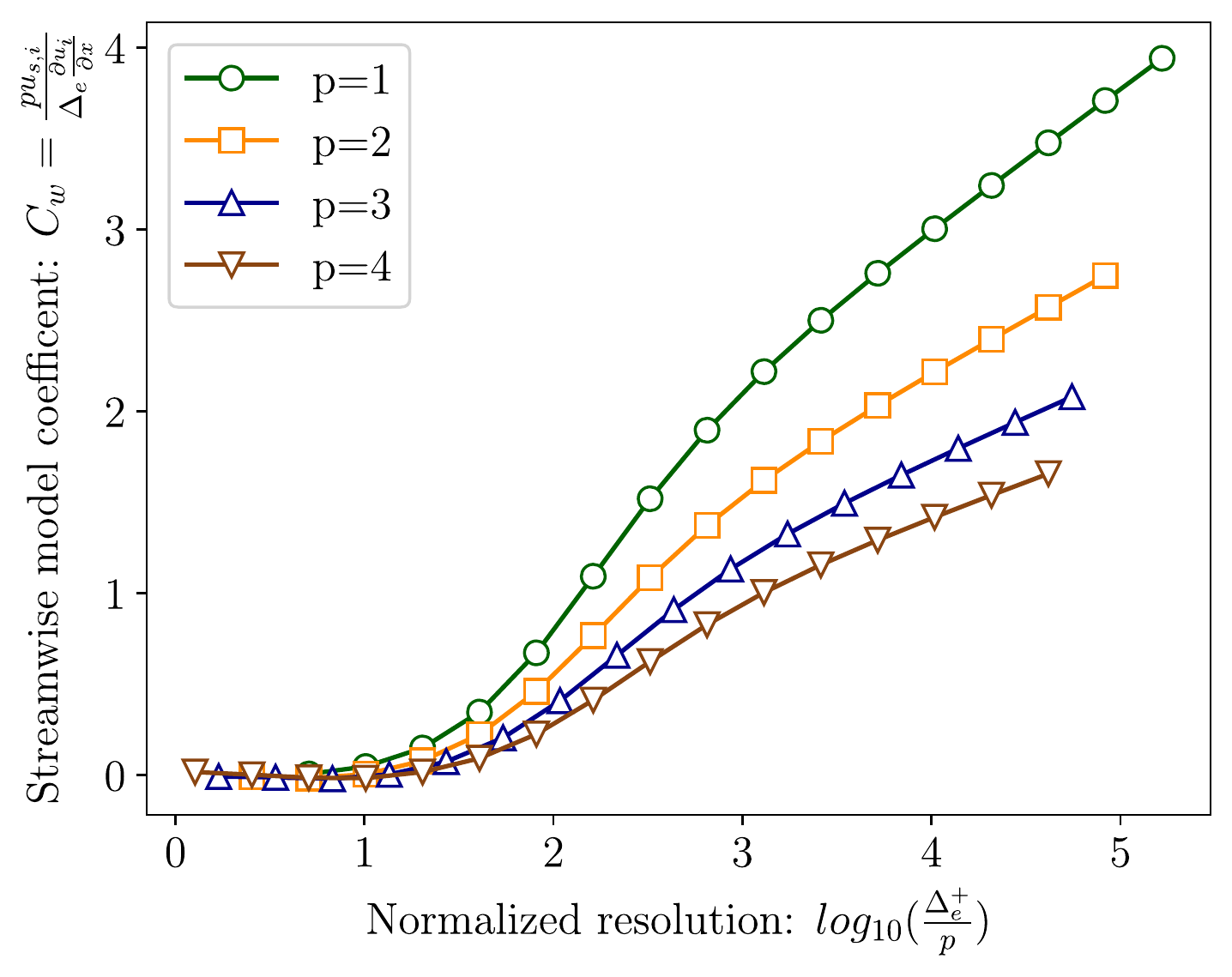}
	\end{subfigure}
	\begin{subfigure}[b]{0.5\textwidth}
		\centering
		\includegraphics[width=\textwidth]{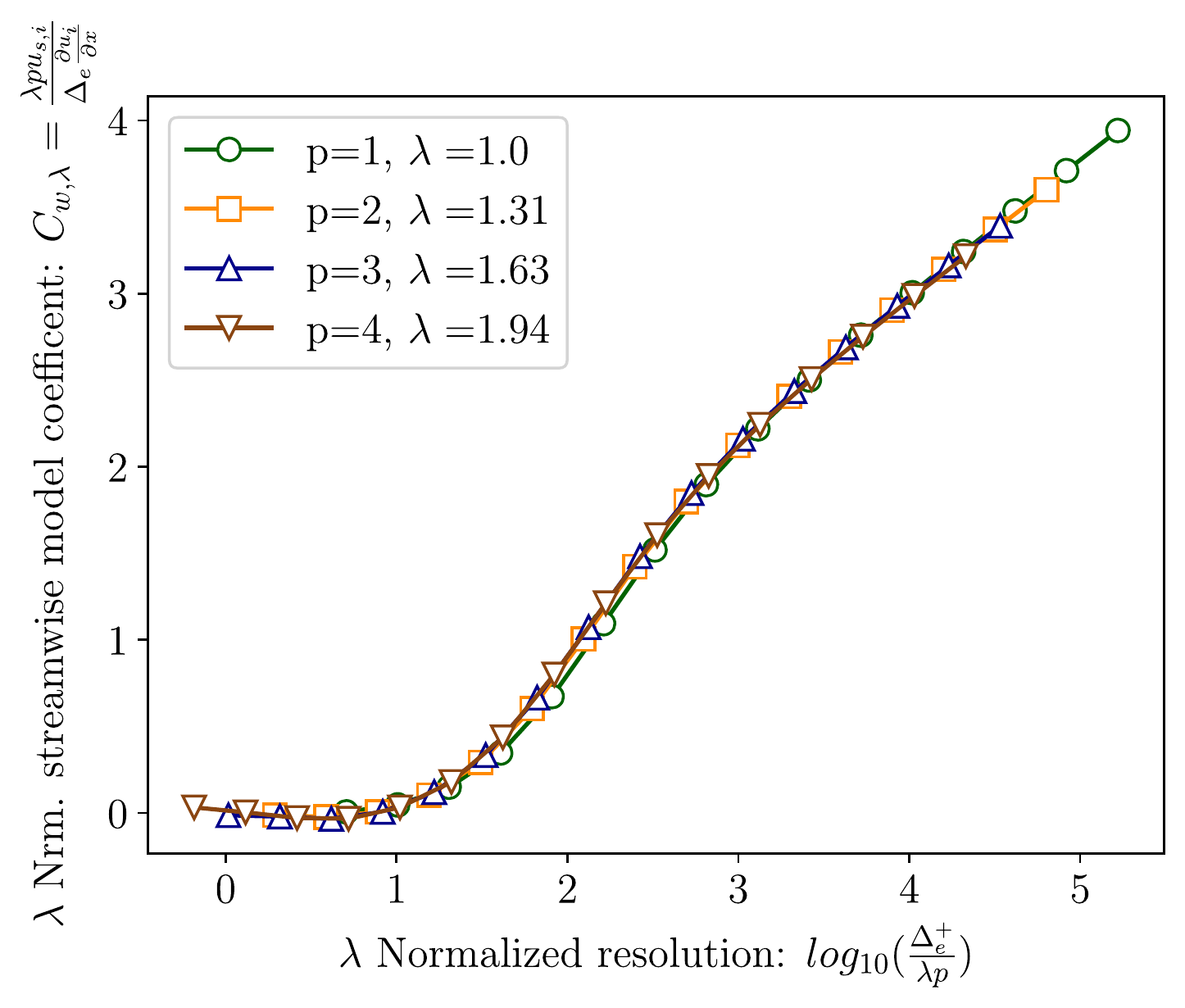}
	\end{subfigure}
		\begin{subfigure}[b]{0.96\textwidth}
		\centering
	\includegraphics[width=\textwidth]{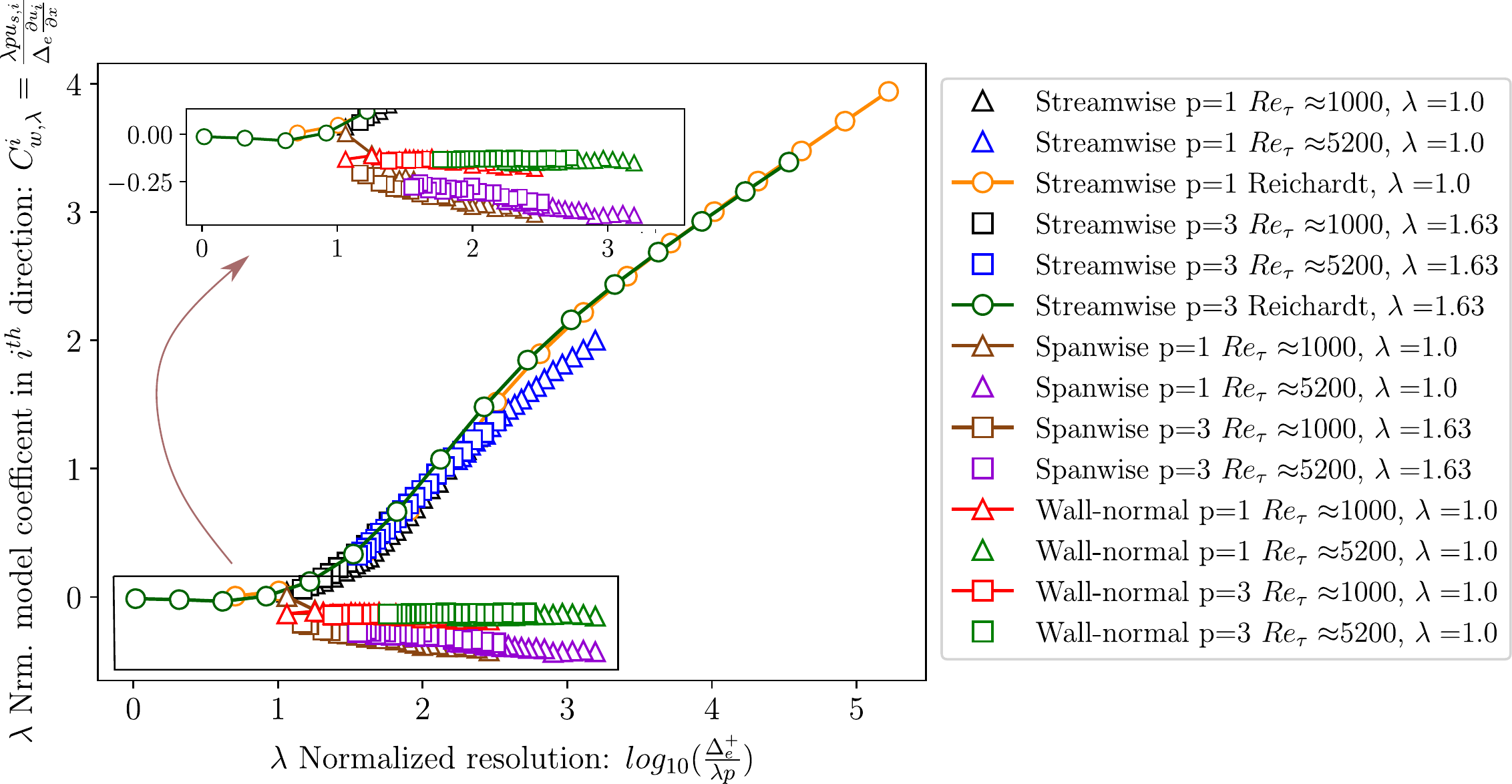}
		\end{subfigure}
\caption{Upper left: $C_w$ computed using Reichardt profile by projecting on different polynomial basis;
Upper right: $\lambda$-Normalized version of Reichardt profiles; 
Bottom: $\lambda$-normalized $C_w$ i.e. $C_{w,\lambda}$ computed by 3D projection of DNS on different polynomial spaces compared to 1-D projection of Reichardt profile. }
\label{cwreichardt}
\end{figure}

 Before investigating the slip velocity due to the full 3-D projection of the DNS solution, it is important to consider the contribution from the mean stream-wise velocity profile itself. As a first step, we will apply 1-D projection to the Reichardt profile \citep{reichardt1951vollstandige} which describes the mean profile in the inner layer. Since the mean solution is invariant in the stream-wise and span-wise direction for a channel, the 3-D projection is reduced to a 1-D projection in the wall-normal direction only. Figure \ref{cwreichardt}(upper left) shows the estimates for $C_w$ obtained from the Reichardt profile for different
orders of projection. It can be observed that the $C_w$ profiles for different orders are distinct even after normalisation of the element size $\Delta_e$ by $p$ to obtain $\Delta = \frac{\Delta_e}{p}$. 

{\color{black} Inspired by the Smagorinsky model, which consists of a model constant $C_s$ that pre-multiplies the grid-size in the final model form, we introduce a new model constant $\lambda$ which in its inverted form i.e. $\frac{1}{\lambda}$ pre-multiplies the grid-size in our proposed model.} Figure \ref{cwreichardt} (upper right) shows the estimated valued for $C_{w,\lambda}$ for different polynomial orders along with the $\lambda$ values for which all the curves collapse to the $p=1$ curve with $\lambda = 1$. As a result, it is possible to learn just one curve and parameterize it with an additional factor $\lambda$ to obtain the $C_w$ curves for different cases i.e.
$C_w = {C_{w,\lambda}}/{\lambda} = g_{1}\left(\Delta^+/\lambda\right)/{\lambda},$
where  $C_{w,\lambda}= \frac{\lambda p u_s}{\Delta_e \frac{\partial u_h}{\partial n}}$. It is also important to check if similar relations also hold true for the 3-D projected solution. 

Figure \ref{cwreichardt} (bottom) compares the $C_{w,\lambda}$ curves obtained through the 3-D projection of DNS solutions to that obtained using the Reichardt mean profile for two different projection orders $p=1,3$. {\color{black}To obtain these curves, large variations in the element sizes have been considered. We use element sizes with $\Delta_e \approx  0.011\delta-0.28 \delta$ for projecting the $Re_{\tau} \approx 1000$ data and element sizes with $\Delta_e \approx  0.033\delta-0.30 \delta$ for projecting the $Re_{\tau} \approx 5200$  data. The  effective filter sizes corresponding to these grids can be approximated by normalizing  the element size with $p$  to obtain $\Delta = \frac{\Delta_e}{p}$} . By re-using the $\lambda$ values from Fig. \ref{cwreichardt}(upper right), similar collapse in the $C_{w,\lambda}$ curves was also obtained for the 3-D projection cases for the stream-wise and the span-wise velocity components. For each projection order, results for two different friction Reynolds numbers of $Re_\tau \approx 1000$ and $Re_\tau \approx 5200$ are plotted. The results indicate that for different polynomial orders, the $C_{w,\lambda}$ estimates for different $Re_{\tau}$ at a particular $\Delta^+$ are same, suggesting that $C_{w,\lambda}$ is a universal function of $\Delta^+$. Further, the $C_{w,\lambda}$ values obtained through 1-D projections of the mean profile are already good approximations to that obtained though the 3-D projections of the DNS solution at moderate resolutions. However, at higher $\Delta^+$, there appears to be a minor discrepancy between the two profiles in the form of a constant shift. The $C_{w,\lambda}$ for the span-wise velocity component is found to be negative and has a slight $\Delta^+$ dependence. Similar to the stream-wise velocity, the $C_{w,\lambda}$ curves for different $Re_\tau$ suggest a $\Delta^+$ dependence in the span-wise direction as well. The $C_w$ in the wall normal direction is approximately zero, and does not depend on the mesh resolution. This suggests that the wall normal slip can be set to zero without the loss of any generalisability. {\color{black} This also suggests that the large-scales, which are typically resolved in a WMLES simulation, can only slide along the wall but cannot penetrate it}. Finally, to obtain a single model form that works for different projection orders, we re-introduce the $\lambda$ factor. In the next section, we will use the insights gained in this section to improve the performance of the slip wall model by \cite{bae2019dynamic,bose2014dynamic} on the channel flow problem.

%\begin{figure}
%    \centering
%    \includegraphics[width=1.00\textwidth]{./slip_DG.pdf}
%    \caption{$C_w$ computed using Reichardt profile by projecting on different polynomial %basis.}
%    \label{cwreichardt}
%\end{figure}

\section{Towards accurate slip-wall models.}\label{sec:slipwall_online}
{\color{black} While the state-of-the-art dynamic slip-wall model by \cite{bae2019dynamic} is found to be better in comparison to the case with no wall-model, it is found to be lacking in accuracy when compared to the traditional WMLES approach. In addition, the slip-wall model has been reported to suffer from instability issues when used   with certain  high-order methods \cite{carton2017assessment}.  Thus, there is a need to improve the stability and performance of slip-wall models on canonical turbulent flow problems before it can be confidently used in more complex flows. Indeed, it is recognized that one disadvantage of the traditional approach is that unlike the dynamic slip-wall model, it requires a priori specification of tunable coefficients. The authors are of the opinion, however, that tunable coefficients should not be used as a reason to replace the traditional WMLES approach which has been shown to perform well across a wider range of problems.  To this end, we try to use data from existing WMLES simulations and our optimal projection techniques to improve the performance of existing slip-wall models to the level of traditional WMLES  for the channel flow problem.  

As we observed in section 3, if the grid resolution is sufficiently coarse, a slip velocity is present at the wall. Hence, it is expected that even the solution from the traditional wall-stress based WMLES  will have a slip-velocity at the wall. Given the excellent performance of the traditional WMLES approach for the channel flow problem, it is also expected that the universal relationship given in figure \ref{cwreichardt} should also hold true for the traditional WMLES approach.  

Traditional WMLES solutions were  computed  using a DG solver with  $p=3$ discretization on different meshes using two different sub-grid models: (i.) a constant coefficient Smagorinsky model with $C_s = 0.12$; (ii.) \cite{VREMEN} model.  Figure \ref{dgdns_rescale2} shows the comparison of $C_{w,\lambda}$ computed using 1-D projection of the Reichardt profile to that computed using the solutions obtained using the traditional WMLES approach.  To compute  $C_{w}$ for a traditional WMLES solution, the solution and its wall normal-gradients are evaluated at the wall to obtain the slip velocity and the pre-multiplied wall-normal gradient. Finally, a least-square fit is performed to obtain a single value of $C_w$. While computing $C_w$, the size of the element $\Delta_e$ is required. However, for all the traditional wall-model cases, the size of the element varies in each direction unlike the grids used for projection of DNS.  As a first attempt, $\Delta_e$ is taken to be the size of the element in the wall-normal direction. Finally, an optimal value of $\lambda$ is found such that the curves collapse asymptotically. By changing the value of  $\lambda$ only the slope of the asymptotic part of the $C_{w,\lambda}$ curve can be changed.  However, when the slope of the $C_{w,\lambda}$ curve in the asymptotic part was made parallel to the $C_{w,\lambda}$ curve obtained for the Reichardt profile by projecting on the $p=1$ basis functions, the intercepts were also found to match. This can be seen in figure \ref{dgdns_rescale2} where the profiles appear identical at large resolutions. However,  small discrepancies exist near the lower resolution limit (i.e. the wall-resolved LES limit), suggesting that either the traditional WMLES approach is inaccurate or the sub-grid model is not accurate. Figure \ref{dgdns_rescale2} also suggests that a universal slip-wall model form exists irrespective of the sub-grid model or the numerical scheme as long as  $\lambda$ is known . 
 
Even if  $\lambda$ is known prior to the simulation or is dynamically determined, a model for $C_{w,\lambda}$ which takes as input the normalized grid-size $\frac{\Delta^+_e}{p \lambda}$ is not useful. This is because the value of $\Delta^+$ is not known unless the wall stress is also known. One option is to use the traditional wall model to obtain the friction velocity $u_{\tau}$ to compute $\Delta^+$ \cite{whitmore2021large}. A better choice would be to represent the slip-wall model coefficient $C_{w,\lambda}$ as a function of the mean slip-velocity $<u_s>$ based Reynolds number i.e. $Re_{slip} = \frac{<u_s> \Delta}{p \lambda \nu}$, as a consequence of which the wall-stress will no longer be required to predict $C_{w,\lambda}$. Figure \ref{dgdns_rescale3} shows the $\lambda$-normalized slip-wall model coefficient  $C_{w,\lambda}$ as a function of the slip-velocity (mean stream-wise) based Reynolds number. As can be observed in figure \ref{dgdns_rescale3}, a universality in the model form similar to the  curves in figure \ref{dgdns_rescale2} also exists in the case when the slip-Reynolds number is used as a feature in place of the normalized grid-size. In addition, the curves were found to collapse to the $p=1$ Reichardt curve for exactly the same value of $\lambda$ used in the case of $C_{w,\lambda}$ vs.  $\Delta^+$. In addition to the plots for $C_{w,\lambda}$ for the various traditional approach obtained using various sub-grid models, a model fit is also provided in the figure \ref{dgdns_rescale2}. This fit can be used as a model to specify $C_{w}$ at the wall as a function of the slip-wall Reynolds number once $\lambda$ is known. 

\begin{figure}
    \centering
    \includegraphics[width=1.0\textwidth]{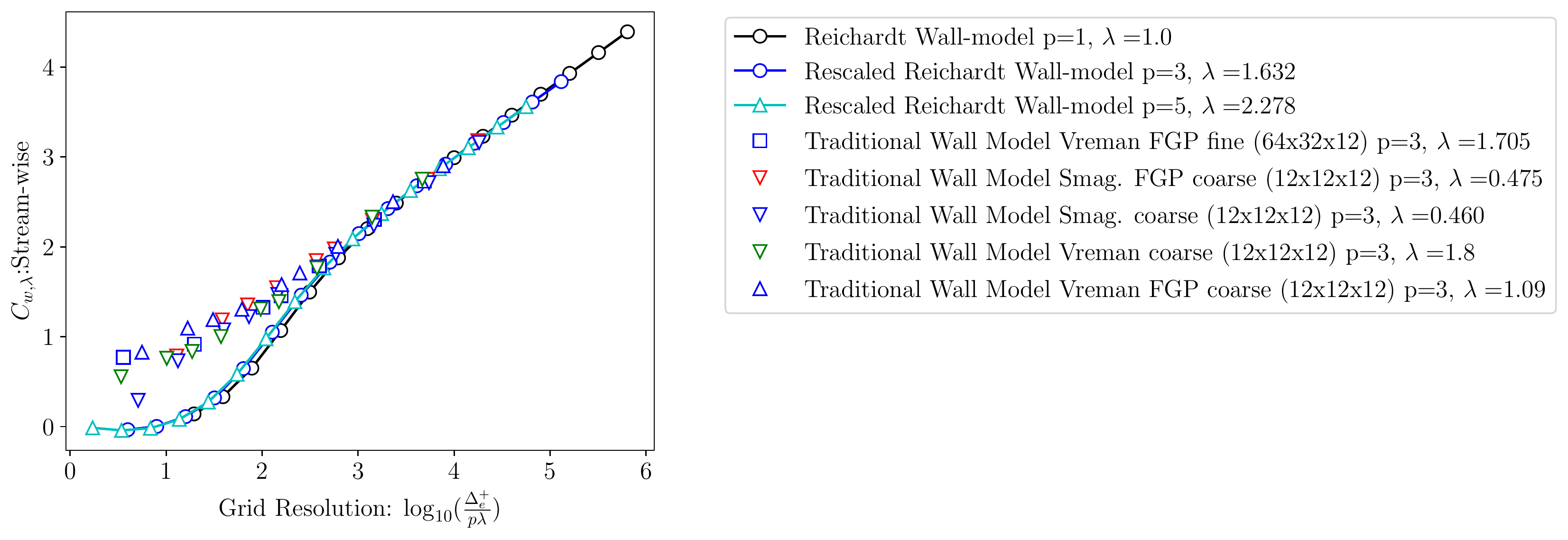}
    \caption{$\lambda$-normalized $C_w$ vs. normalized grid-size $\frac{\Delta^+_e}{p \lambda}$. The $\lambda$-normalized $C_w$ are computed by 1-D projection of Reichardt profile and compared to the same obtained using the traditional WMLES solution. The plots marked by "FGP" use only the explicit sub-grid models inside the first element and gradually change to implicit LES outside the first element.}
    \label{dgdns_rescale2}
\end{figure}

\begin{figure}
    \centering
    \includegraphics[width=1.0\textwidth]{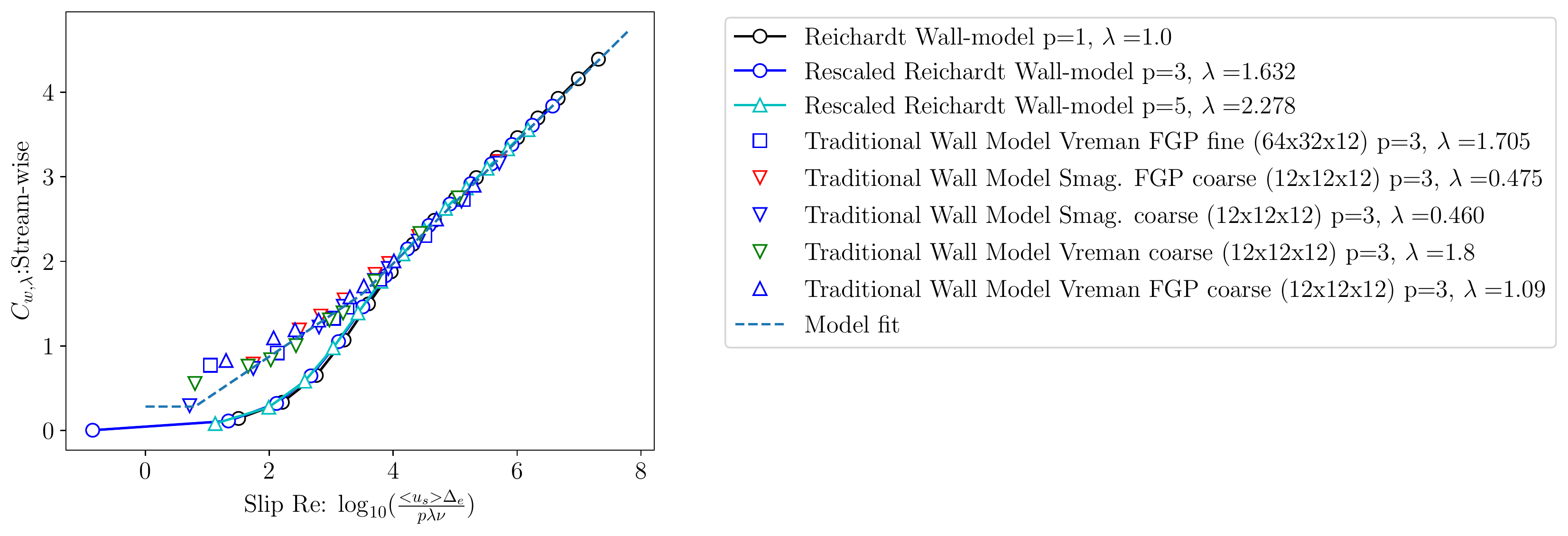}
    \caption{$\lambda$-normalized $C_w$ vs. the slip-velocity $u_s$ based Reynolds number $\frac{<u_s> \Delta}{p \lambda \nu}$. The $\lambda$-normalized $C_w$ are computed by 1-D projection of Reichardt profile and compared to the same obtained using the traditional WMLES solution. The plots marked by "FGP" use the explicit sub-grid models inside the first element only and gradually change to implicit LES outside the first element.}
    \label{dgdns_rescale3}
\end{figure}

As a first step, we will apply the $C_w$ computed using the traditional approach and apply it as a slip-boundary condition to check if the traditional WMLES results can be recreated with the slip boundary condition. At this stage, we are applying the same value of $C_w$ for the stream-wise and the span-wise components. In this implementation, it is assumed that the there is no transpiration i.e. no flow through the wall. To apply the slip-wall boundary condition we first use the slip velocity components $u_{s,i}$ at the wall to compute the wall normal derivatives of the velocity components $u_{h,i}$ as follows: 

\begin{equation}
    \frac{\partial u_{h,i}}{\partial n} = \frac{u_{s,i}}{\Delta C_w},
\end{equation}
and finally compute the wall stress at any location using the following formula:
\begin{equation}
\tau_{w,i}=\nu \left.\frac{\partial {u}_{h,i}}{\partial n}\right|_{w} -\left.\tau_{i,n}^{S G S}\right|_{w}.
\end{equation}
 Hence, contribution of the mean wall stress is only present from the viscous and the sub-grid stresses. In addition to the $C_w$ obtained by post-processing the traditional approach solutions, the $C_w$ computed using the slip-Reynolds number based model are also used. The value of $\lambda$, required for implementing the slip-Reynolds number based approach is obtained from the traditional method. The slip-Reynolds number based model does not require the specification of different $C_w$'s for each $Re_{\tau}$ case, however, requires one $\lambda$ which remains constant across all the cases with different $Re_{\tau}$. Figure \ref{slip_recreate} shows the stream-wise mean velocity profiles, the R.M.S of different velocity components and the Reynolds shear stress profiles at different friction Reynolds numbers. The vertical dashed lines show the location of the 1st, 2nd and 3rd off-wall grid points. For the traditional wall model, the wall-stress is computed using the velocity components at the 3rd off-wall grid point. The slip wall model does not require any such exchange location. It is clear from figure \ref{slip_recreate} that when the correct $\lambda$ is used, the Reynolds number dependence is captured accurately and the statistics obtained using the slip-wall model are identical to the traditional wall model. 

\begin{figure}
 \centering
 \includegraphics[width=1.00\textwidth]{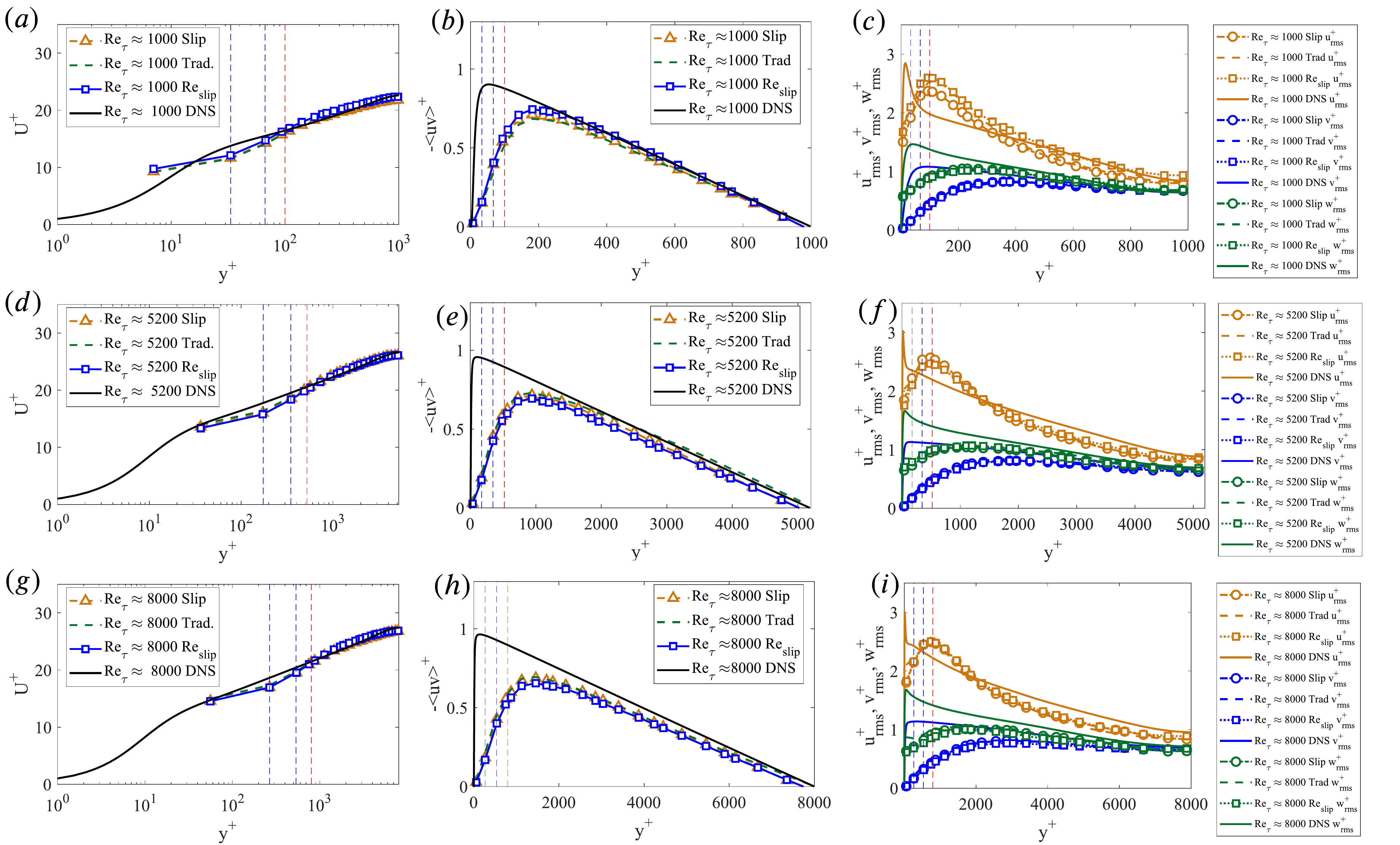}
\caption{Comparison of the first-order and second-order statistics using the traditional method (Trad.), by re-using the slip-wall model with $C_w$ computed from the traditional WMLES solution (Slip), and with the $C_w$ computed using the slip Reynolds number formulation ($Re_{slip}$) at different friction Reynolds number. The vertical dashed lines show the locations of the first, second and third grid points. For the traditional wall-model, velocity is sampled at the third off-wall grid point.}
\label{slip_recreate}
\end{figure}

The previous tests presented in figure \ref{slip_recreate} showed that the model is able to capture the Reynolds number dependence on a single grid. The next step is to change the grid-resolution and check if similar results also hold true for the new grid. Before performing numerical experiments with our proposed slip-wall model, an a priori study could be performed by using the results from the traditional WMLES solutions. Two different meshes are now used with $12 \times 12 \times 12$ and $18 \times 18 \times 12$ elements, respectively. The resolution in the wall normal direction is kept the same, whereas, the resolution in the span-wise and stream-wise case are smaller in case of the $18 \times 18 \times 12$ mesh because the size of the channel is kept constant. 

Figure \ref{dgdns_rescale4} shows  the $\lambda$-normalized $C_w$ obtained for different resolutions for two different types of normalizations. Different normalizations are used because the effective $\Delta$ is not known in the case where the element is not cubic. The plots marked by "WN" and "VOL" use the wall-normal grid-spacing and the cube root of the cell volume as $\Delta_e$, respectively. It can be observed that when the wall-normal grid-spacing is used as $\Delta_e$, the $\lambda$ values required for the two different resolutions are different. This suggests that if the wall-normal grid resolution is used for $\Delta_e$, our proposed slip-wall model will not generalise to a different grid for the same $\lambda$ value. On the other hand, when the cell volume was used for $\Delta_e$, the $\lambda$ values required to ensure that both the curves collapse was found to be same. This suggests that, for the resolutions considered here, the cube root of the cell volume is an ideal candidate for $\Delta_e$ to ensure that the slip-wall model generalises to a new grid for the same value of $\lambda$.  Hence, the proposed model will require the specification of the model constant $\lambda$ and it is expected to work on different grids and Reynolds numbers. Figure \ref{dgdns_rescale4} also shows plot of  $C_{w,\lambda}$ at two other resolutions of $12 \times 12 \times 16$ and $32 \times 16 \times 12$. While constructing the model-fit these resolutions have not been considered. These plots for these specific resolutions will be later used to explain the success of the slip-wall model on these unseen resolutions. 

Figures \ref{slip_predict} shows the stream-wise mean velocity profiles, the R.M.S of different velocity components and the Reynolds shear stress profiles at different friction Reynolds numbers on two different meshes with $12 \times 12 \times 16$  and $32 \times 16 \times 12$ elements, respectively, that are not part of the data used for fitting the model for $C_{w,\lambda}$. Clearly, the model not only captures the effect of $Re_{\tau}$ but also generalises to a new resolution. The performance of the proposed slip-wall model is comparable to the traditional wall model which is a major improvement over the dynamic slip-wall model proposed by \cite{bae2019dynamic}. The results indicate that the proposed wall-model is able to work reasonably well even at considerably different resolutions. The excellent performance of the slip-wall model can be also explained by computing $C_{w,\lambda}$ using the traditional WMLES solutions on these grids. The $C_{w,\lambda}$ values estimated using the traditional WMLES solutions from two different meshes with $12 \times 12 \times 16$  and $32 \times 16 \times 12$ elements, respectively, are plotted in figure \ref{dgdns_rescale4}. The accurate prediction of $C_{w,\lambda}$ by the model fit explains the excellent predictive performance of our slip-wall model . }

\begin{figure}
    \centering 
    \includegraphics[width=1.0\textwidth]{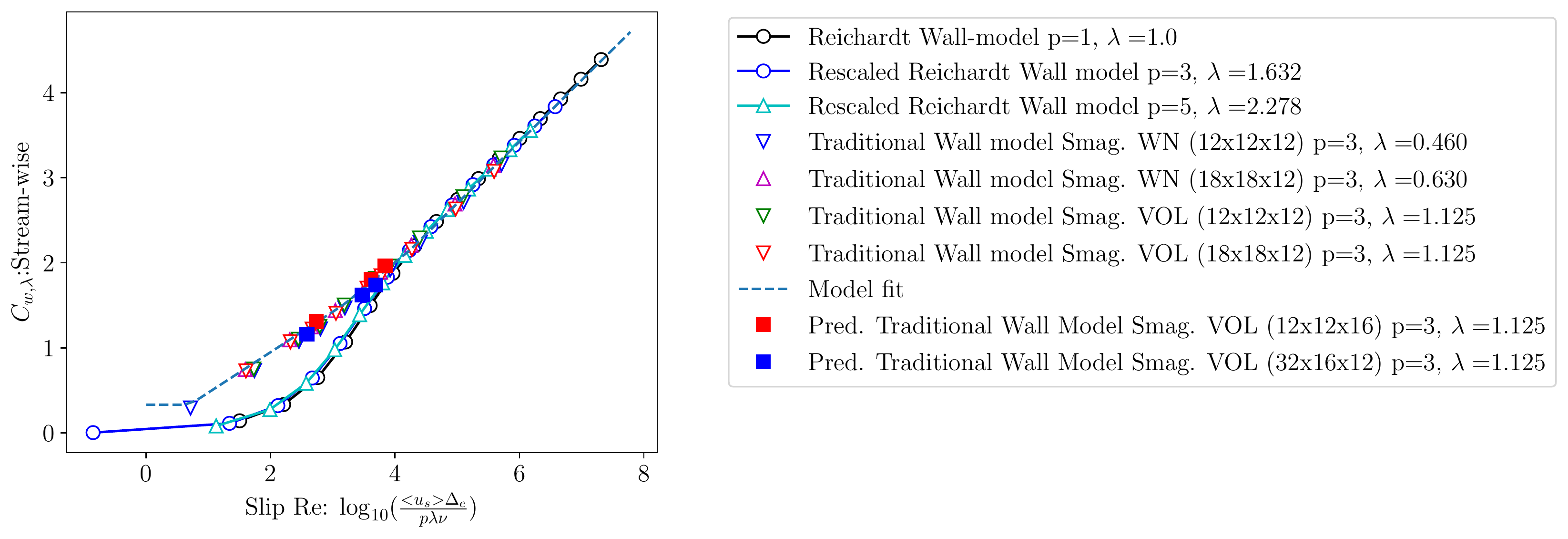}
    \caption{$\lambda$-normalized $C_w$ vs. the mean slip-velocity $<u_s>$ based Reynolds number $\frac{<u_s> \Delta}{p \lambda \nu}$. The $\lambda$-normalized $C_w$ are computed by 1-D projection of Reichardt profile and compared to the same obtained using the traditional WMLES solution. The plots marked by "WN" and "VOL" use the wall-normal grid-spacing and the cube root of the cell volume for specification of $\Delta_e$, respectively.}
    \label{dgdns_rescale4}
\end{figure}

\begin{figure}
 \centering
 \includegraphics[width=1.00\textwidth]{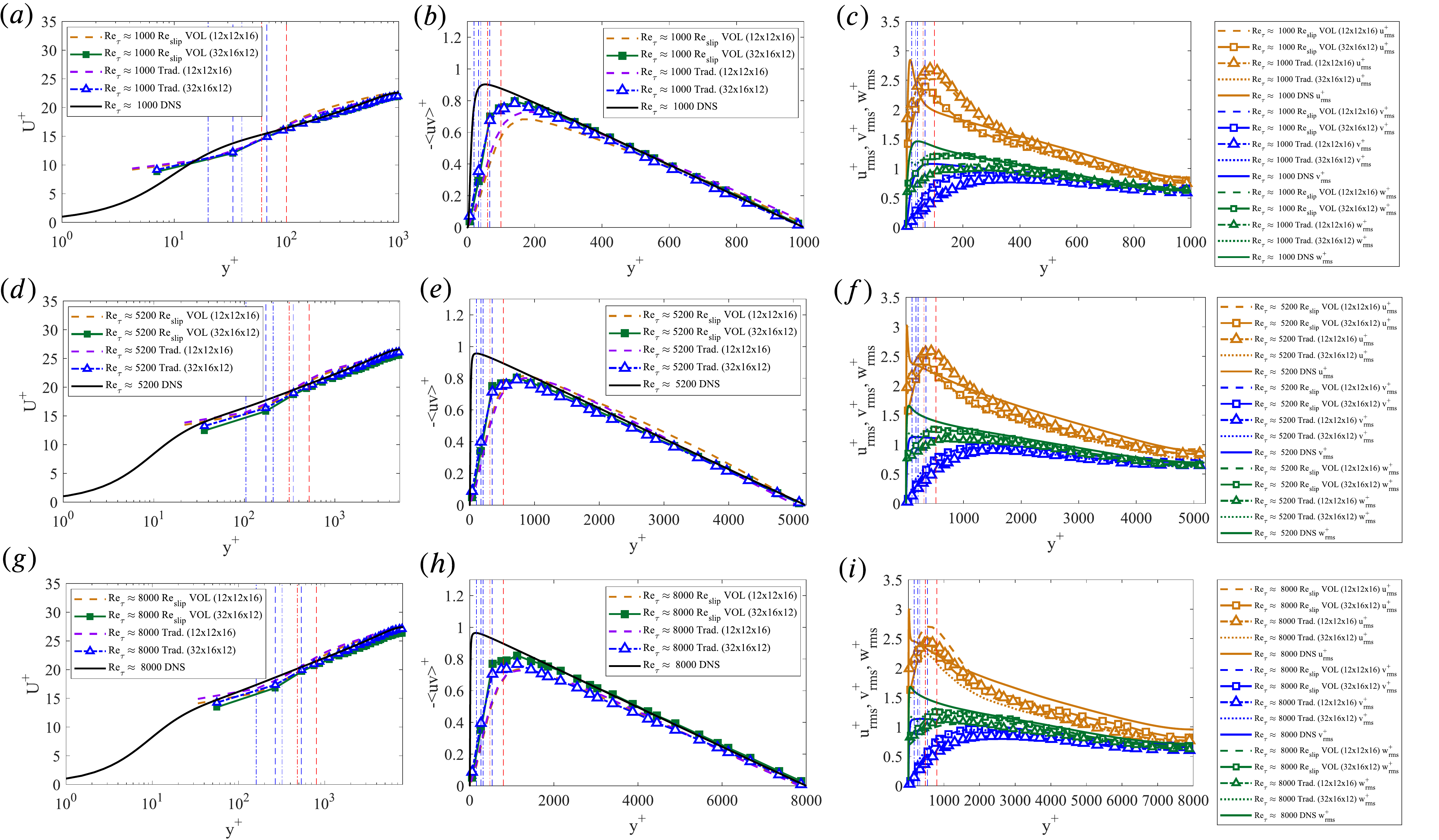}
\caption{Comparison of the first-order and second-order statistics obtained using the traditional method (Trad.) and the proposed slip-wall model at different friction Reynolds numbers. The solution is computed on two different meshes with $12 \times 12 \times 16$ and $32 \times 16 \times 12$ elements, respectively, that is not part of the data used for fitting the model. The vertical dash-dotted and dashed lines show the locations of first, second and third grid points for the meshes with $12 \times 12 \times 16$ and $32 \times 16 \times 12$ elements, respectively. For the traditional wall-model, velocity is sampled at the third off-wall grid point.}
\label{slip_predict}
\end{figure}
\section{Perspectives on Improved slip-wall models.}
 Slip-based wall models~\cite{bose2014dynamic,bae2019dynamic,whitmore2021large} allow for easy implementation of the wall boundary condition without the need to sample velocity components at a few grid points away from the wall and also allows the possibility to model flow separation.  {\color{black} The dynamic slip-wall model proposed by ~\cite{bae2019dynamic} shows excellent performance on the zero-pressure gradient flat plate case, albeit for a narrow range of $Re_\theta$. However, this model requires improvements when compared to traditional wall models for the equilibrium channel flow case at similar resolutions. In section 4,  we provided a priori results on the model form for $C_w$ for equilibrium channel flows.  In Section 5, we used some of the insights obtained from section 4 to improve the performance of the existing slip-wall model to atleast the traditional WMLES level. Although the performance of our proposed slip wall model was found to be acceptable,  more insights from section 4 can be used to further improve the accuracy of both the proposed and the existing slip wall models.} To this end, we outline the following ingredients for the construction of a more generalizable slip-based wall model forms :    
\begin{enumerate}
\item The slip model coefficients can be different in the stream-wise, span-wise and the wall-normal directions as observed in figure \ref{cwreichardt}(bottom). In a more complex 3-D case, the choice of stream-wise, span-wise and the wall-normal direction is a bit ambiguous. However, the mean-flow can be used to identify these directions. However, this needs to be iteratively done since the mean flow can itself change when changing these directions. Another approach is to use the flow direction at the first off-wall grid point, similar to how the traditional wall models are implemented. {\color{black} In Section 5, we used the same  $C_w$ for all directions. The effect of using different $C_w$ for different velocity components on our proposed model is a topic of further research .}
\item If a dynamic modeling procedure is performed to obtain $C_w$, the value of $C_w$ cannot be assumed to be same at the original grid and the test filtered grid. Figure \ref{cwreichardt}(bottom) shows that  $C_w$ changes when the resolution is changed from $\Delta^+$ to the test filtered grid resolution $2\Delta^+$. In addition to $C_w$ being not constant across different grid levels, there is a dependence on the wall-units. This dependence is generally not considered in the existing slip-wall model forms. However, this dependence is present in  traditional wall models which are found to perform excellently for equilibrium wall-bounded flow cases.{  \color{black}In Section 5, we were able to improve the performance of the dynamic slip-wall model of \cite{bae2019dynamic} by just augmenting the model form without performing any dynamic procedure.}
\item {\color{black}The discrepancy in figures \ref{dgdns_rescale2} and \ref{dgdns_rescale3} between the $C_{w,\lambda}$ curves obtained by the optimal projection of the Reichardt profile, and that obtained using the solutions of the traditional WMLES approach suggests that the current WMLES approaches are sub-optimal due to the presence of wall-modeling and sub-grid modeling errors. This also suggests that there is a lot of scope for improvements, and our optimal projection framework can be used to assess the WMLES performance of different combination of sub-grid models and wall models.}
\item The final comment is on the choice of the parameter that should be used to performing the dynamic procedure. We saw in section 4 that the value of $\lambda$ effectively captures the effect of the order of projection and hence the numerical method. The corresponding function $g_1$ is fairly universal for different orders. Hence, it is imperative that the dynamic modeling be performed on $\lambda$ rather than $C_w$. {\color{black} The model form for $g_1$ can be empirically obtained from DNS data, Reichardt profile or from the solution of an existing model such as the wall-stress based WMLES models. We further observed in section 5 that if the cube root of the cell volume is used for $\Delta_e$, $\lambda$ remains fairly constant across different resolutions and Reynolds numbers for a given sub-grid model. A dynamic model that determines $\lambda$ without requiring the solutions from the traditional model is a topic of further research.}
\end{enumerate}
 
{\color{black} In this work, two different model forms for $C_{w,\lambda}$ were proposed as shown in figure \ref{dgdns_rescale2} and figure \ref{dgdns_rescale3}. The first model uses $\Delta^+$ as a feature, whereas, the second model uses the  Reynolds number based on the slip-velocity as a feature. The implementation of the first model is slightly more complex because the proposed expression for $C_w$ is a function of two parameters: $\lambda$ and $\Delta^+$. Assuming that the grid size $\Delta$ is known, to compute $\Delta^+$ from $\Delta$, an estimate of the average wall stress $<\tau_w>$ is required: 
\begin{equation}
\left\langle\tau_{w}\right\rangle=v\left\langle\left.\frac{\partial {u}_{h,1}}{\partial y}\right|_{w}\right\rangle-\left\langle\left.{u}_{h,1} {u}_{h,2}\right|_{w}\right\rangle-\left\langle\left.\tau_{12}^{S G S}\right|_{w}\right\rangle.
\end{equation}
This average quantity influences the slip velocities through $C_w$ which in turn affects the average itself. As discussed earlier, an alternate approach is to use the equilibrium wall-profile to obtain $\tau_w$ as done in \cite{whitmore2021large}. However, this requires the sampling of  the velocity fields from the off-wall grid points which makes the implementation of slip-wall models as cumbersome as the traditional wall model. An alternate approach is to use  the  Reynolds number based on the slip-velocity as described in section 5. In this work, the optimal estimates of $C_w$ were obtained from the DNS solution by projecting on uniform elements of different sizes. However, our projection framework by using anisotropic elements also allows us to study the effect of the grid aspect ratios. As discussed in section 5, one approach to account for mesh anisotropy is to replace $\Delta_e$ with an effective grid size such as the cube root of the cell volume. However, it is advisable to include the aspect ratio in the model form as well to ensure optimal performance across different types of meshes.  Finally, the present model form has been derived from the channel flow data and its accuracy in the spatially developing flows such as the flat plate or in the separated flow regions has not been assessed. Given the excellent performance of the traditional wall models on the flat-plat cases we expect similar performance from our proposed model, however, this is a topic of further research.   }

\section{Conclusion.}
%A finite element-based filtering approach was proposed. 
{\color{black}
The projection-based scale-separation approach is an essential part of the variational multiscale method and uses the grid effectively as a filter. It is applicable to cases where the filter length is anisotopic, varies in space or filtering needs to performed on an unstructured grid. These filter properties were found to be essential for a priori assessment of existing coarse-grained methods for wall-bounded turbulent flows, where the grids can be highly anisotropic and vary in size along a particular direction. 

An a priori assessment of the optimal solutions at three different limits: the wall-resolved LES, the hybrid RANS-LES and the WMLES limit, was performed by projecting DNS on different grids suitable for these scale-resolving approaches. For each of these cases, while projecting the DNS on to the coarse-space, weak imposition of the boundary condition was made by not enforcing no-slip boundary conditions at the boundary nodes. In the wall-resolved LES limit, the mean velocity was found to be well-resolved, no-slip was naturally satisfied and the turbulent stresses were well represented. In the hybrid RANS-LES limit, which was obtained by coarsening the wall-resolved LES mesh in the span-wise and stream-wise directions, the mean velocity was well-resolved and the no-slip boundary condition was naturally satisfied. However, the turbulent stresses were found to be well represented only at the center of the channel and under-represented in the near-wall region where sufficient resolution was not present. In the WMLES limit, which is obtained by further coarsening the hybrid RANS-LES grid in the wall normal direction, the mean profile is no longer represented accurately near the wall and a slip-velocity is obtained. The turbulent stresses in WMLES are relatively well-represented at the center of the channel compared to the near-wall region. In the near-wall region, the stream-wise and the span-wise velocity fluctuations were found to be non-zero at the wall, whereas, the the resolved wall-normal fluctuations and the turbulent shear-stress were found to be under-represented. All these trends were found to be consistent with existing solutions in the literature suggesting that the present framework can be utilized to assess, augment and calibrate existing methods.  

The ability to obtain slip-velocity directly from 3-D projection of DNS on coarse near-wall meshes enabled further assessment of the existing slip-wall based wall-models. As a first step, estimates of the slip-wall model coefficient $C_w$ were obtained from the mean velocity profile in the inner-layer through 1-D projections of the Reichardt profile. The $C_w$ estimates from the mean-profile were found to be strongly dependent on the order of projection suggesting that the numerical method has considerable impact on the optimal value of $C_w$. In addition to this, the resolution for a given slip velocity and projection order was found to scale with the wall units. To make modeling more tractable, we introduced an extra resolution normalizer $\lambda$ to express the effect of projection order through a single coefficient, similar in scope to the Smagorinsky model coefficient $C_s$. When this analysis was extended to 3-D, similar dependence on the polynomial order $p$ on $C_w$ was found for the stream-wise and the span-wise velocity components. However, on re-introduction of resolution normalizer $\lambda$ and reusing the $\lambda$ values corresponding to the 1-D projections, similar collapse in the $C_{w,\lambda}$ values was also observed for the 3-D case. The value of $C_{w,\lambda}$ was also found to be different for the stream-wise, span-wise and the wall-normal velocity components.

The ultimate goal of  any a-priori analysis is to improve the model performance in  a posteriori calculations.  As a first step towards better slip based wall models, the performance of existing slip-based wall models was compared to  traditional WMLES for channel flows. To establish an equivalence between the two methods,  $C_{w,\lambda}$ curves were evaluated using the solution of the traditional WMLES approach and compared with the curves obtained for the Reichardt profile. The $C_{w,\lambda}$ curves for the traditional WMLES solutions were found to  be identical to those obtained using the Reichardt profile at high $\frac{\Delta_e^+}{\lambda p}$.  However, at low  $\frac{\Delta_e^+}{\lambda p}$, the $C_{w,\lambda}$, the curves were found to differ suggesting the presence of sub-grid modeling and wall-modelling errors in the solution. To reduce the implementation challenges associated with using $\frac{\Delta_e^+}{\lambda p}$ as a feature, a slip Reynolds number-based ($Re_{slip}$) feature was introduced.  Finally, by choosing $\Delta_e$ to be the cube root of the cell volume and re-using $\lambda$ from the traditional WMLES solution, a model form was constructed by fitting the $C_{w,\lambda}$ vs.  $Re_{slip}$ curve. The resulting model was shown to generalize to different resolutions, element aspect ratios and Reynolds numbers in a posteriori simulations.}

\section*{Acknowledgement}

This research was funded by NASA under the project "Scale-resolving turbulence simulations through adaptive high-order discretizations and data-enabled model refinements", grant number 80NSSC18M0149 (Technical monitor: Dr. Gary Coleman). We  acknowledge Prof. Krzysztof Fidkowski for  valuable discussions.

\section*{Declaration of interests.}

The authors report no conflict of interest.

\appendix
\section{Numerical computation of $L_2$-projection.}\label{appendixA}
{\color{black} The problem of finding an $L_2$ projection is equivalent to the problem of finding a $u_h\in {\mathcal{V}_h}$ such that   
\begin{equation}
(u_h,w_h) = (u,w_h)  \quad \forall {w_h} \in {{\mathcal{V}_h}}.
\label{proj_app}
\end{equation}
The first step is to determine the coarse space $\mathcal{V}_h$. The coarse space should be low-dimensional in comparison to the original solution to ensure that the projection operation acts as a filter. The low dimensionality of the coarse-space can be ensured by using lesser number of grid points or modes. There are many choices for the coarse space (for e.g, the Fourier basis functions, the global Chebyshev polynomial basis functions and the piece-wise polynomial basis functions). Once the coarse space is fixed, the coarse solution can be written as a linear combination of the basis functions as follows:
\begin{equation}
u_h = \mathbf{w}^{T}_h \mathbf{a}_h
\label{proj_app2}
\end{equation}
where $\mathbf{w}^{T}_h$ is a vector of coarse-scale basis functions spanning the coarse-space and $\mathbf{a}_h$ is vector containing the corresponding basis coefficients. Substituting equation (\ref{proj_app2}) in equation \ref{proj_app} we obtain:
\begin{equation}
 \mathbf{M} \mathbf{a}_h  = \mathbf{r},
\label{proj_app3}
\end{equation}
where the mass matrix $\mathbf{M}$ and the right hand side vector $\mathbf{r}$ is given by,
\begin{equation}
 \mathbf{M}  = \int \mathbf{w}_h \mathbf{w}^{T}_h d\Omega, \textrm{ and }
\mathbf{r} = (u,\mathbf{w}_h).
\end{equation}
The coarse-scale basis coefficients are obtained as $
 \mathbf{a}_h  = \mathbf{M}^{-1} \mathbf{r}.$
The mass matrix $\mathbf{M}$ is local (block diagonal) when DG basis functions are used. In the case of CG basis functions, the mass matrix has to be assembled by adding contributions from individual element mass matrices. The computation of $\mathbf{M}$ is not as expensive as compared to the right hand side vector $\mathbf{r}$, especially when $u$ is high dimensional. The elements of the matrix $\mathbf{M}$ can be precisely computed using a Gauss quadrature rule appropriate for the order of the polynomial used to define the coarse-space. The computation of $(u,\mathbf{w}_h)$, however, needs special care because it requires the computation of the inner-product of a high dimensional solution $u$ with the coarse basis functions $w_h$ as shown in figure \ref{dgproj_RHS}. The high dimensional solution $u$ can come from a finite difference, finite volume, spectral or finite element simulation.

A general approach to compute the elements of the right hand side vector $\mathbf{r}$ is by using numerical integration. As can be observed in figure \ref{dgproj_RHS}, the solution obtained after multiplication of the coarse basis functions $w_{h,i}$ with $u$ still contains high-dimensional features and requires a fine-grid for numerical integration. The grid on which $u$ exists is assumed to be sufficiently fine for performing the numerical integration. In case the projected solution depends on the order of numerical integration or the size of the numerical integration grid, the solution can be injected on a more finer grid to perform the numerical integration. Once the integration grid is set, the Trapezoidal rule or the Simpson's formula can be applied to compute the integral over the $uw_{h,i}$ fields to obtain the right hand side vector $\mathbf{r}$. 

\begin{figure}
    \centering
    \includegraphics[width=1.0\textwidth]{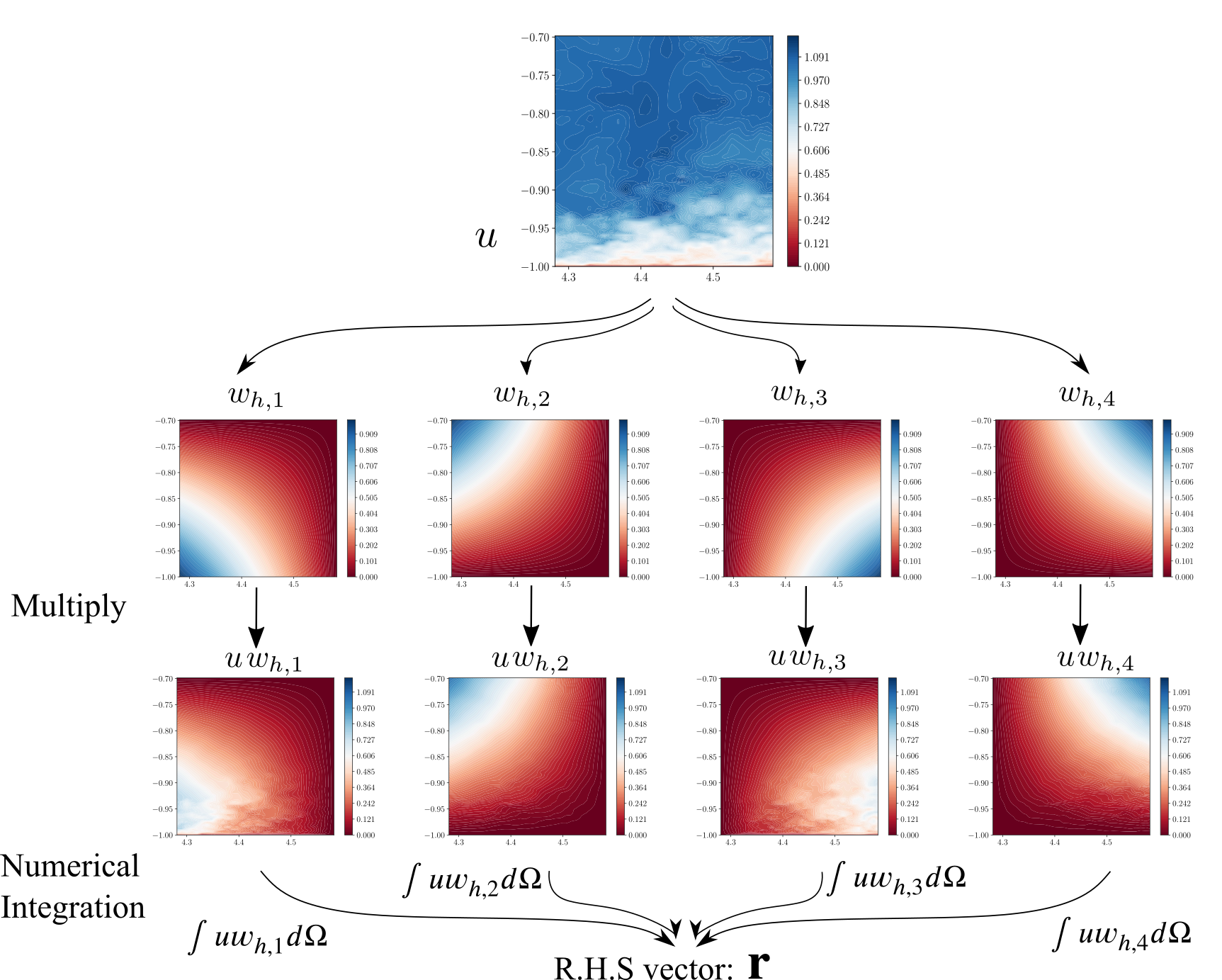}
    \caption{The high dimensional solution $u$ is multiplied with DG coarse-scale basis $w_{h,i}$ to obtain $uw_{h,i}$. The right hand side $\mathbf{r}$ is finally computed by evaluating $\int uw_{h,i} d\Omega$ for all basis function $w_{h,i}$ spanning the coarse space.}
    \label{dgproj_RHS}
\end{figure}
}

\bibliographystyle{jfm}
\bibliography{jfm}

\begin{thebibliography}{63}
\expandafter\ifx\csname natexlab\endcsname\relax\def\natexlab#1{#1}\fi
\def\au#1{#1} \def\ed#1{#1} \def\yr#1{#1}\def\at#1{#1}\def\jt#1{\textit{#1}}
  \def\bt#1{#1}\def\bvol#1{\textbf{#1}} \def\vol#1{#1} \def\pg#1{#1}
  \def\publ#1{#1}\def\arxiv#1{#1}\def\org#1{#1}\def\st#1{\textit{#1}}

\bibitem[Bae {\em et~al.\/}(2019)Bae, Lozano-Dur{\'a}n, Bose \&
  Moin]{bae2019dynamic}
{\sc \au{Bae, Hyunji~Jane}, \au{Lozano-Dur{\'a}n, Adri{\'a}n}, \au{Bose,
  Sanjeeb~T} \& \au{Moin, Parviz}} \yr{2019}  \at{Dynamic slip wall model for
  large-eddy simulation}.  \jt{Journal of fluid mechanics}  \bvol{859},
  \pg{400--432}.

\bibitem[Bazilevs {\em et~al.\/}(2007)Bazilevs, Calo, Cottrell, Hughes, Reali
  \& Scovazzi]{VMS3}
{\sc \au{Bazilevs, Y}, \au{Calo, VM}, \au{Cottrell, JA}, \au{Hughes, TJR},
  \au{Reali, A} \& \au{Scovazzi, G}} \yr{2007}  \at{Variational multiscale
  residual-based turbulence modeling for large eddy simulation of
  incompressible flows}.  \jt{Computer Methods in Applied Mechanics and
  Engineering}  \bvol{197}~(1-4),  \pg{173--201}.

\bibitem[Beck {\em et~al.\/}(2019)Beck, Flad \& Munz]{beck2019deep}
{\sc \au{Beck, Andrea}, \au{Flad, David} \& \au{Munz, Claus-Dieter}} \yr{2019}
  \at{Deep neural networks for data-driven les closure models}.  \jt{Journal of
  Computational Physics}  \bvol{398},  \pg{108910}.

\bibitem[Bose \& Moin(2014)]{bose2014dynamic}
{\sc \au{Bose, Sanjeeb~T} \& \au{Moin, Parviz}} \yr{2014}  \at{A dynamic slip
  boundary condition for wall-modeled large-eddy simulation}.  \jt{Physics of
  Fluids}  \bvol{26}~(1),  \pg{015104}.

\bibitem[Bou-Zeid {\em et~al.\/}(2008)Bou-Zeid, Vercauteren, Parlange \&
  Meneveau]{bou2008scale}
{\sc \au{Bou-Zeid, Elie}, \au{Vercauteren, Nikki}, \au{Parlange, Marc~B} \&
  \au{Meneveau, Charles}} \yr{2008}  \at{Scale dependence of subgrid-scale
  model coefficients: an a priori study}.  \jt{Physics of Fluids}
  \bvol{20}~(11),  \pg{115106}.

\bibitem[Chung \& Freund(2022)]{chung2022optimization}
{\sc \au{Chung, Seung~Whan} \& \au{Freund, Jonathan~B}} \yr{2022}  \at{An
  optimization method for chaotic turbulent flow}.  \jt{Journal of
  Computational Physics}  \bvol{457},  \pg{111077}.

\bibitem[Codina(2002)]{OSS}
{\sc \au{Codina, Ramon}} \yr{2002}  \at{Stabilized finite element approximation
  of transient incompressible flows using orthogonal subscales}.  \jt{Computer
  Methods in Applied Mechanics and Engineering}  \bvol{191}~(39-40),
  \pg{4295--4321}.

\bibitem[Codina {\em et~al.\/}(2007)Codina, Principe, Guasch \& Badia]{OSS2}
{\sc \au{Codina, Ramon}, \au{Principe, Javier}, \au{Guasch, Oriol} \&
  \au{Badia, Santiago}} \yr{2007}  \at{Time dependent subscales in the
  stabilized finite element approximation of incompressible flow problems}.
  \jt{Computer Methods in Applied Mechanics and Engineering}
  \bvol{196}~(21-24),  \pg{2413--2430}.

\bibitem[Del~Alamo {\em et~al.\/}(2004)Del~Alamo, Jim{\'e}nez, Zandonade \&
  Moser]{del2004scaling}
{\sc \au{Del~Alamo, Juan~C}, \au{Jim{\'e}nez, Javier}, \au{Zandonade, Paulo} \&
  \au{Moser, Robert~D}} \yr{2004}  \at{Scaling of the energy spectra of
  turbulent channels}.  \jt{Journal of Fluid Mechanics}  \bvol{500},
  \pg{135--144}.

\bibitem[Duraisamy(2021)]{duraisamy2021perspectives}
{\sc \au{Duraisamy, Karthik}} \yr{2021}  \at{Perspectives on machine
  learning-augmented reynolds-averaged and large eddy simulation models of
  turbulence}.  \jt{Physical Review Fluids}  \bvol{6}~(5),  \pg{050504}.

\bibitem[Duraisamy {\em et~al.\/}(2019)Duraisamy, Iaccarino \&
  Xiao]{duraisamy2019turbulence}
{\sc \au{Duraisamy, Karthik}, \au{Iaccarino, Gianluca} \& \au{Xiao, Heng}}
  \yr{2019}  \at{Turbulence modeling in the age of data}.  \jt{Annual Review of
  Fluid Mechanics}  \bvol{51},  \pg{357--377}.

\bibitem[Friess \& Davidson(2020)]{friess2020formulation}
{\sc \au{Friess, Christophe} \& \au{Davidson, Lars}} \yr{2020}  \at{A
  formulation of pans capable of mimicking iddes}.  \jt{International Journal
  of Heat and Fluid Flow}  \bvol{86},  \pg{108666}.

\bibitem[Gamahara \& Hattori(2017)]{gamahara2017searching}
{\sc \au{Gamahara, Masataka} \& \au{Hattori, Yuji}} \yr{2017}  \at{Searching
  for turbulence models by artificial neural network}.  \jt{Physical Review
  Fluids}  \bvol{2}~(5),  \pg{054604}.

\bibitem[Germano(1986)]{germano1986differential}
{\sc \au{Germano, Massimo}} \yr{1986}  \at{Differential filters of elliptic
  type}.  \jt{The Physics of fluids}  \bvol{29}~(6),  \pg{1757--1758}.

\bibitem[Germano {\em et~al.\/}(1991)Germano, Piomelli, Moin \& Cabot]{DSM}
{\sc \au{Germano, Massimo}, \au{Piomelli, Ugo}, \au{Moin, Parviz} \& \au{Cabot,
  William~H}} \yr{1991}  \at{A dynamic subgrid-scale eddy viscosity model}.
  \jt{Physics of Fluids A: Fluid Dynamics}  \bvol{3}~(7),  \pg{1760--1765}.

\bibitem[Girimaji \& Abdol-Hamid(2005)]{girimaji2005partially}
{\sc \au{Girimaji, Sharath} \& \au{Abdol-Hamid, Khaled}} \yr{2005}
  Partially-averaged navier stokes model for turbulence: Implementation and
  validation.  \bt{In {\em 43rd AIAA Aerospace Sciences Meeting and
  Exhibit\/}},  \pg{p. 502}.

\bibitem[Goc {\em et~al.\/}(2020)Goc, Bose \& Moin]{goc2020wall}
{\sc \au{Goc, Konrad}, \au{Bose, Sanjeeb} \& \au{Moin, Parviz}} \yr{2020}
  Wall-modeled large eddy simulation of an aircraft in landing configuration.
  \bt{In {\em AIAA Aviation 2020 Forum\/}},  \pg{p. 3002}.

\bibitem[Goc {\em et~al.\/}(2021)Goc, Lehmkuhl, Park, Bose \&
  Moin]{goc2021large}
{\sc \au{Goc, Konrad~A}, \au{Lehmkuhl, Oriol}, \au{Park, George~Ilhwan},
  \au{Bose, Sanjeeb~T} \& \au{Moin, Parviz}} \yr{2021}  \at{Large eddy
  simulation of aircraft at affordable cost: a milestone in computational fluid
  dynamics}.  \jt{Flow}  \bvol{1}.

\bibitem[Gopalan {\em et~al.\/}(2013)Gopalan, Heinz \&
  St{\"o}llinger]{gopalan2013unified}
{\sc \au{Gopalan, Harish}, \au{Heinz, Stefan} \& \au{St{\"o}llinger,
  Michael~K}} \yr{2013}  \at{A unified rans--les model: Computational
  development, accuracy and cost}.  \jt{Journal of Computational Physics}
  \bvol{249},  \pg{249--274}.

\bibitem[Gravemeier {\em et~al.\/}(2010)Gravemeier, Gee, Kronbichler \&
  Wall]{NLVMS}
{\sc \au{Gravemeier, Volker}, \au{Gee, Michael~W}, \au{Kronbichler, Martin} \&
  \au{Wall, Wolfgang~A}} \yr{2010}  \at{An algebraic variational
  multiscale--multigrid method for large eddy simulation of turbulent flow}.
  \jt{Computer Methods in Applied Mechanics and Engineering}
  \bvol{199}~(13-16),  \pg{853--864}.

\bibitem[Heinz(2007)]{heinz2007unified}
{\sc \au{Heinz, Stefan}} \yr{2007}  \at{Unified turbulence models for les and
  rans, fdf and pdf simulations}.  \jt{Theoretical and Computational Fluid
  Dynamics}  \bvol{21}~(2),  \pg{99--118}.

\bibitem[Hoyas \& Jim{\'e}nez(2006)]{hoyas2006scaling}
{\sc \au{Hoyas, Sergio} \& \au{Jim{\'e}nez, Javier}} \yr{2006}  \at{Scaling of
  the velocity fluctuations in turbulent channels up to re $\tau$= 2003}.
  \jt{Physics of fluids}  \bvol{18}~(1),  \pg{011702}.

\bibitem[Hughes {\em et~al.\/}(1998)Hughes, Feij{\'o}o, Mazzei \&
  Quincy]{hughes1998variational}
{\sc \au{Hughes, Thomas~JR}, \au{Feij{\'o}o, Gonzalo~R}, \au{Mazzei, Luca} \&
  \au{Quincy, Jean-Baptiste}} \yr{1998}  \at{The variational multiscale
  method—a paradigm for computational mechanics}.  \jt{Computer methods in
  applied mechanics and engineering}  \bvol{166}~(1-2),  \pg{3--24}.

\bibitem[Iyer \& Malik(2020)]{iyer2020wall}
{\sc \au{Iyer, Prahladh~S} \& \au{Malik, Mujeeb~R}} \yr{2020} Wall-modeled les
  of the nasa juncture flow experiment.  \bt{In {\em AIAA Scitech 2020
  Forum\/}},  \pg{p. 1307}.

\bibitem[Kawai \& Larsson(2012)]{kawai2012wall}
{\sc \au{Kawai, Soshi} \& \au{Larsson, Johan}} \yr{2012}  \at{Wall-modeling in
  large eddy simulation: Length scales, grid resolution, and accuracy}.
  \jt{Physics of Fluids}  \bvol{24}~(1),  \pg{015105}.

\bibitem[Kiris {\em et~al.\/}(2022)Kiris, Ghate, Duensing, Browne, Housman,
  Stich, Kenway, Dos Santos~Fernandes \& Machado]{kiris2022high}
{\sc \au{Kiris, Cetin~C}, \au{Ghate, Aditya~S}, \au{Duensing, Jared~C},
  \au{Browne, Oliver~M}, \au{Housman, Jeffrey~A}, \au{Stich, Gerrit-Daniel},
  \au{Kenway, Gaetan}, \au{Dos Santos~Fernandes, Luis~M} \& \au{Machado,
  Leonardo~M}} \yr{2022} High-lift common research model: Rans, hrles, and
  wmles perspectives for clmax prediction using lava.  \bt{In {\em AIAA SCITECH
  2022 Forum\/}},  \pg{p. 1554}.

\bibitem[Krank \& Wall(2016)]{krank2016new}
{\sc \au{Krank, Benjamin} \& \au{Wall, Wolfgang~A}} \yr{2016}  \at{A new
  approach to wall modeling in les of incompressible flow via function
  enrichment}.  \jt{Journal of Computational Physics}  \bvol{316},
  \pg{94--116}.

\bibitem[Langford \& Moser(1999)]{optimal}
{\sc \au{Langford, Jacob~A} \& \au{Moser, Robert~D}} \yr{1999}  \at{Optimal
  {LES} formulations for isotropic turbulence}.  \jt{Journal of fluid
  mechanics}  \bvol{398},  \pg{321--346}.

\bibitem[Lee \& Moser(2015)]{lee2015direct}
{\sc \au{Lee, Myoungkyu} \& \au{Moser, Robert~D}} \yr{2015}  \at{Direct
  numerical simulation of turbulent channel flow up to}.  \jt{Journal of fluid
  mechanics}  \bvol{774},  \pg{395--415}.

\bibitem[Li {\em et~al.\/}(2008)Li, Perlman, Wan, Yang, Meneveau, Burns, Chen,
  Szalay \& Eyink]{li2008public}
{\sc \au{Li, Yi}, \au{Perlman, Eric}, \au{Wan, Minping}, \au{Yang, Yunke},
  \au{Meneveau, Charles}, \au{Burns, Randal}, \au{Chen, Shiyi}, \au{Szalay,
  Alexander} \& \au{Eyink, Gregory}} \yr{2008}  \at{A public turbulence
  database cluster and applications to study lagrangian evolution of velocity
  increments in turbulence}.  \jt{Journal of Turbulence} ~(9),  \pg{N31}.

\bibitem[Lozano-Duran {\em et~al.\/}(2020)Lozano-Duran, Bose \&
  Moin]{lozano2020prediction}
{\sc \au{Lozano-Duran, Adrian}, \au{Bose, Sanjeeb~T} \& \au{Moin, Parviz}}
  \yr{2020} Prediction of trailing edge separation on the nasa juncture flow
  using wall-modeled les.  \bt{In {\em AIAA Scitech 2020 Forum\/}},  \pg{p.
  1776}.

\bibitem[Masud \& Calderer(2011)]{NLVMS2}
{\sc \au{Masud, Arif} \& \au{Calderer, Ramon}} \yr{2011}  \at{A variational
  multiscale method for incompressible turbulent flows: {Bubble} functions and
  fine scale fields}.  \jt{Computer Methods in Applied Mechanics and
  Engineering}  \bvol{200}~(33-36),  \pg{2577--2593}.

\bibitem[Maulik \& San(2017)]{maulik2017neural}
{\sc \au{Maulik, Romit} \& \au{San, Omer}} \yr{2017}  \at{A neural network
  approach for the blind deconvolution of turbulent flows}.  \jt{Journal of
  Fluid Mechanics}  \bvol{831},  \pg{151--181}.

\bibitem[Maulik {\em et~al.\/}(2019)Maulik, San, Jacob \& Crick]{maulik2019sub}
{\sc \au{Maulik, Romit}, \au{San, Omer}, \au{Jacob, Jamey~D} \& \au{Crick,
  Christopher}} \yr{2019}  \at{Sub-grid scale model classification and blending
  through deep learning}.  \jt{Journal of Fluid Mechanics}  \bvol{870},
  \pg{784--812}.

\bibitem[Maulik {\em et~al.\/}(2018)Maulik, San, Rasheed \&
  Vedula]{maulik2018data}
{\sc \au{Maulik, Romit}, \au{San, Omer}, \au{Rasheed, Adil} \& \au{Vedula,
  Prakash}} \yr{2018}  \at{Data-driven deconvolution for large eddy simulations
  of {Kraichnan} turbulence}.  \jt{Physics of Fluids}  \bvol{30}~(12),
  \pg{125109}.

\bibitem[Meneveau \& Katz(2000)]{meneveau2000scale}
{\sc \au{Meneveau, Charles} \& \au{Katz, Joseph}} \yr{2000}
  \at{Scale-invariance and turbulence models for large-eddy simulation}.
  \jt{Annual Review of Fluid Mechanics}  \bvol{32}~(1),  \pg{1--32}.

\bibitem[Meneveau {\em et~al.\/}(1996)Meneveau, Lund \& Cabot]{DSM2}
{\sc \au{Meneveau, Charles}, \au{Lund, Thomas~S} \& \au{Cabot, William~H}}
  \yr{1996}  \at{A lagrangian dynamic subgrid-scale model of turbulence}.
  \jt{Journal of fluid mechanics}  \bvol{319},  \pg{353--385}.

\bibitem[Najafi-Yazdi {\em et~al.\/}(2015)Najafi-Yazdi, Najafi-Yazdi \&
  Mongeau]{najafi2015high}
{\sc \au{Najafi-Yazdi, Alireza}, \au{Najafi-Yazdi, Mostafa} \& \au{Mongeau,
  Luc}} \yr{2015}  \at{A high resolution differential filter for large eddy
  simulation: Toward explicit filtering on unstructured grids}.  \jt{Journal of
  Computational Physics}  \bvol{292},  \pg{272--286}.

\bibitem[Nicoud \& Ducros(1999)]{WALE}
{\sc \au{Nicoud, Franck} \& \au{Ducros, Fr{\'e}d{\'e}ric}} \yr{1999}
  \at{Subgrid-scale stress modelling based on the square of the velocity
  gradient tensor}.  \jt{Flow, turbulence and Combustion}  \bvol{62}~(3),
  \pg{183--200}.

\bibitem[Nicoud {\em et~al.\/}(2011)Nicoud, Toda, Cabrit, Bose \& Lee]{SIGMA}
{\sc \au{Nicoud, Franck}, \au{Toda, Hubert~Baya}, \au{Cabrit, Olivier},
  \au{Bose, Sanjeeb} \& \au{Lee, Jungil}} \yr{2011}  \at{Using singular values
  to build a subgrid-scale model for large eddy simulations}.  \jt{Physics of
  Fluids}  \bvol{23}~(8),  \pg{085106}.

\bibitem[Parish \& Duraisamy(2017)]{MZVMS}
{\sc \au{Parish, Eric~J} \& \au{Duraisamy, Karthik}} \yr{2017}  \at{A unified
  framework for multiscale modeling using the {Mori-Zwanzig} formalism and the
  variational multiscale method}.  \jt{arXiv preprint arXiv:1712.09669} .

\bibitem[Park \& Moin(2016)]{park2016wall}
{\sc \au{Park, GI} \& \au{Moin, P}} \yr{2016}  \at{Wall-modeled les: Recent
  applications to complex flows}.  \jt{Annual Research Briefs}  \pg{pp.
  39--50}.

\bibitem[Perlman {\em et~al.\/}(2007)Perlman, Burns, Li \&
  Meneveau]{perlman2007data}
{\sc \au{Perlman, Eric}, \au{Burns, Randal}, \au{Li, Yi} \& \au{Meneveau,
  Charles}} \yr{2007} Data exploration of turbulence simulations using a
  database cluster.  \bt{In {\em Proceedings of the 2007 ACM/IEEE Conference on
  Supercomputing\/}},  \pg{pp. 1--11}.

\bibitem[Piomelli \& Balaras(2002)]{piomelli2002wall}
{\sc \au{Piomelli, Ugo} \& \au{Balaras, Elias}} \yr{2002}  \at{Wall-layer
  models for large-eddy simulations}.  \jt{Annual review of fluid mechanics}
  \bvol{34}~(1),  \pg{349--374}.

\bibitem[Pope(2000)]{pope2000turbulent}
{\sc \au{Pope, Stephen~B}} \yr{2000} {\em Turbulent flows\/}.  \publ{Cambridge
  university press}.

\bibitem[Pradhan \& Duraisamy(2021)]{pradhan2021variational}
{\sc \au{Pradhan, Aniruddhe} \& \au{Duraisamy, Karthik}} \yr{2021}
  \at{Variational multi-scale super-resolution: A data-driven approach for
  reconstruction and predictive modeling of unresolved physics}.  \jt{arXiv
  preprint arXiv:2101.09839} .

\bibitem[Reichardt(1951)]{reichardt1951vollstandige}
{\sc \au{Reichardt, Hans}} \yr{1951}  \at{Vollst{\"a}ndige darstellung der
  turbulenten geschwindigkeitsverteilung in glatten leitungen}.
  \jt{ZAMM-Journal of Applied Mathematics and Mechanics/Zeitschrift f{\"u}r
  Angewandte Mathematik und Mechanik}  \bvol{31}~(7),  \pg{208--219}.

\bibitem[Sarghini {\em et~al.\/}(2003)Sarghini, De~Felice \&
  Santini]{sarghini2003neural}
{\sc \au{Sarghini, F}, \au{De~Felice, G} \& \au{Santini, S}} \yr{2003}
  \at{Neural networks based subgrid scale modeling in large eddy simulations}.
  \jt{Computers \& fluids}  \bvol{32}~(1),  \pg{97--108}.

\bibitem[Shur {\em et~al.\/}(1999)Shur, Spalart, Strelets \&
  Travin]{shur1999detached}
{\sc \au{Shur, M}, \au{Spalart, PR}, \au{Strelets, M} \& \au{Travin, A}}
  \yr{1999}  \at{Detached-eddy simulation of an airfoil at high angle of
  attack}.  \bt{In {\em Engineering turbulence modelling and experiments 4\/}},
   \pg{pp. 669--678}.  \publ{Elsevier}.

\bibitem[Shur {\em et~al.\/}(2008)Shur, Spalart, Strelets \&
  Travin]{shur2008hybrid}
{\sc \au{Shur, Mikhail~L}, \au{Spalart, Philippe~R}, \au{Strelets, Mikhail~Kh}
  \& \au{Travin, Andrey~K}} \yr{2008}  \at{A hybrid rans-les approach with
  delayed-des and wall-modelled les capabilities}.  \jt{International journal
  of heat and fluid flow}  \bvol{29}~(6),  \pg{1638--1649}.

\bibitem[Spalart(2009)]{spalart2009detached}
{\sc \au{Spalart, Philippe~R}} \yr{2009}  \at{Detached-eddy simulation}.
  \jt{Annual review of fluid mechanics}  \bvol{41},  \pg{181--202}.

\bibitem[Vreman(2004)]{VREMEN}
{\sc \au{Vreman, AW}} \yr{2004}  \at{An eddy-viscosity subgrid-scale model for
  turbulent shear flow: Algebraic theory and applications}.  \jt{Physics of
  fluids}  \bvol{16}~(10),  \pg{3670--3681}.

\bibitem[Vreman {\em et~al.\/}(1995)Vreman, Geurts \&
  Kuerten]{vreman1995priori}
{\sc \au{Vreman, Bert}, \au{Geurts, Bernard} \& \au{Kuerten, Hans}} \yr{1995}
  \at{A priori tests of large eddy simulation of the compressible plane mixing
  layer}.  \jt{Journal of engineering mathematics}  \bvol{29}~(4),
  \pg{299--327}.

\bibitem[Wang {\em et~al.\/}(2020)Wang, Hu \& Zheng]{wang2020comparative}
{\sc \au{Wang, Limin}, \au{Hu, Ruifeng} \& \au{Zheng, Xiaojing}} \yr{2020}
  \at{A comparative study on the large-scale-resolving capability of
  wall-modeled large-eddy simulation}.  \jt{Physics of Fluids}  \bvol{32}~(3),
  \pg{035102}.

\bibitem[Wang {\em et~al.\/}(2018)Wang, Luo, Li, Tan \&
  Fan]{wang2018investigations}
{\sc \au{Wang, Zhuo}, \au{Luo, Kun}, \au{Li, Dong}, \au{Tan, Junhua} \&
  \au{Fan, Jianren}} \yr{2018}  \at{Investigations of data-driven closure for
  subgrid-scale stress in large-eddy simulation}.  \jt{Physics of Fluids}
  \bvol{30}~(12),  \pg{125101}.

\bibitem[Wang \& Oberai(2010)]{VMSE}
{\sc \au{Wang, Z} \& \au{Oberai, AA}} \yr{2010}  \at{Spectral analysis of the
  dissipation of the residual-based variational multiscale method}.
  \jt{Computer Methods in Applied Mechanics and Engineering}
  \bvol{199}~(13-16),  \pg{810--818}.

\bibitem[Whitmore {\em et~al.\/}(2021)Whitmore, Griffin, Bose \&
  Moin]{whitmore2021large}
{\sc \au{Whitmore, MP}, \au{Griffin, KP}, \au{Bose, ST} \& \au{Moin, P}}
  \yr{2021}  \at{Large-eddy simulation of a gaussian bump with slip-wall
  boundary conditions}.  \jt{Center for Turbulence Research Annual Research
  Briefs}  \pg{pp. 45--58}.

\bibitem[Carton~de Wiart \& Murman(2017)]{carton2017assessment}
{\sc \au{Carton~de Wiart, Corentin} \& \au{Murman, Scott~M}} \yr{2017}
  Assessment of wall-modeled les strategies within a discontinuous-galerkin
  spectral-element framework.  \bt{In {\em 55th AIAA Aerospace Sciences
  Meeting\/}},  \pg{p. 1223}.

\bibitem[Xie {\em et~al.\/}(2019{\natexlab{{\em a\/}}})Xie, Li, Ma \&
  Wang]{xie2019modeling}
{\sc \au{Xie, Chenyue}, \au{Li, Ke}, \au{Ma, Chao} \& \au{Wang, Jianchun}}
  \yr{2019{\natexlab{{\em a\/}}}}  \at{Modeling subgrid-scale force and
  divergence of heat flux of compressible isotropic turbulence by artificial
  neural network}.  \jt{Physical Review Fluids}  \bvol{4}~(10),  \pg{104605}.

\bibitem[Xie {\em et~al.\/}(2019{\natexlab{{\em b\/}}})Xie, Wang, Li, Wan \&
  Chen]{xie2019artificial}
{\sc \au{Xie, Chenyue}, \au{Wang, Jianchun}, \au{Li, Hui}, \au{Wan, Minping} \&
  \au{Chen, Shiyi}} \yr{2019{\natexlab{{\em b\/}}}}  \at{Artificial neural
  network mixed model for large eddy simulation of compressible isotropic
  turbulence}.  \jt{Physics of Fluids}  \bvol{31}~(8),  \pg{085112}.

\bibitem[Xie {\em et~al.\/}(2020)Xie, Wang \& Weinan]{xie2020modeling}
{\sc \au{Xie, Chenyue}, \au{Wang, Jianchun} \& \au{Weinan, E}} \yr{2020}
  \at{Modeling subgrid-scale forces by spatial artificial neural networks in
  large eddy simulation of turbulence}.  \jt{Physical Review Fluids}
  \bvol{5}~(5),  \pg{054606}.

\bibitem[Yang \& Griffin(2021)]{yang2021grid}
{\sc \au{Yang, Xiang~IA} \& \au{Griffin, Kevin~P}} \yr{2021}  \at{Grid-point
  and time-step requirements for direct numerical simulation and large-eddy
  simulation}.  \jt{Physics of Fluids}  \bvol{33}~(1),  \pg{015108}.

\bibitem[You \& Moin(2007)]{GDSM}
{\sc \au{You, Donghyun} \& \au{Moin, Parviz}} \yr{2007}  \at{A dynamic
  global-coefficient subgrid-scale eddy-viscosity model for large-eddy
  simulation in complex geometries}.  \jt{Physics of Fluids}  \bvol{19}~(6),
  \pg{065110}.

\end{thebibliography}

\end{document}